\documentclass[aps,prl,twocolumn,superscriptaddress,showpacs,nofootinbib]{revtex4-1}
\usepackage{graphicx,lipsum}
\usepackage{amsmath,amssymb,color}
\usepackage{pgfplots}
\usepackage[export]{adjustbox}
\usepackage[normalem]{ulem}
\usepackage{bm}
\usepackage{pifont}

\definecolor{cyan1}{RGB}{51 153 255}
\definecolor{magenta1}{RGB}{255 0 102}
\definecolor{dark_blue2}{RGB}{0 0 153}
\definecolor{black1}{RGB}{0 0 0}
\definecolor{violet1}{RGB}{153 0 204}
\definecolor{orange1}{RGB}{255 51 0}
\definecolor{dark_red}{RGB}{204 0 0}
\definecolor{dark_green2}{RGB}{0 102 0}
\definecolor{light_yellow}{RGB}{255 255 204}
\definecolor{brown1}{RGB}{102 51 0}
\definecolor{dark_yellow}{RGB}{204 102 0}

\def\orange#1{\textcolor{orange1}{#1}}
\def\cyan#1{\textcolor{cyan1}{#1}}

\def\dblue#1{\textcolor{dark_blue2}{#1}}

\def\violet#1{\textcolor{violet1}{#1}}
\def\dred#1{\textcolor{dark_red}{#1}}
\def\dgreen#1{\textcolor{dark_green2}{#1}}

\def\brown#1{\textcolor{brown1}{#1}}
\def\dyellow#1{\textcolor{dark_yellow}{#1}}

\begin{document}

\title{Crossover from interaction to driven regimes\\
in quantum vortex reconnections}

\author{Luca Galantucci}
\email[]{luca.galantucci@newcastle.ac.uk}
\affiliation{Joint Quantum Centre (JQC) Durham--Newcastle, and
School of Mathematics and Statistics, Newcastle University,
Newcastle upon Tyne, NE1 7RU, United Kingdom}

\author{A.~W.~Baggaley}
\affiliation{Joint Quantum Centre (JQC) Durham--Newcastle, and
School of Mathematics and Statistics, Newcastle University,
Newcastle upon Tyne, NE1 7RU, United Kingdom}

\author{N.~G.~Parker}
\affiliation{Joint Quantum Centre (JQC) Durham--Newcastle, and
School of Mathematics and Statistics, Newcastle University,
Newcastle upon Tyne, NE1 7RU, United Kingdom}

\author{C.~F. Barenghi}
\affiliation{Joint Quantum Centre (JQC) Durham--Newcastle, and
School of Mathematics and Statistics, Newcastle University,
Newcastle upon Tyne, NE1 7RU, United Kingdom}

\date{\today}

\begin{abstract}
Reconnections of coherent filamentary structures
play a key role in the dynamics of fluids,
redistributing energy and helicity among the length scales,
triggering dissipative effects and inducing fine-scale mixing.
Unlike ordinary (classical) fluids where vorticity is a continuous field,
in superfluid helium and in atomic Bose-Einstein condensates (BECs)
vorticity takes the form of isolated quantised vortex lines, which are
conceptually easier to study.
New experimental techniques now allow visualisation of individual
vortex reconnections in helium and condensates.
It has long being suspected that reconnections obey universal
laws, particularly a universal scaling with time of the minimum distance between
vortices $\delta$.  Here we perform a comprehensive analysis of this scaling across a range of scenarios relevant to superfluid helium and trapped condensates, 
combining our own numerical simulations with the previous results in the literature.  We reveal that the scaling exhibit two distinct 
fundamental regimes: a $\delta \sim t^{1/2}$ scaling arising from the mutual interaction of the reconnecting strands and a $\delta \sim t$ scaling when extrinsic factors drive the  individual vortices.
\end{abstract}


\maketitle


\section*{Reconnections in classical and quantum systems}  

\begin{figure}[htbp]
\centering
\includegraphics[width=0.9\columnwidth]{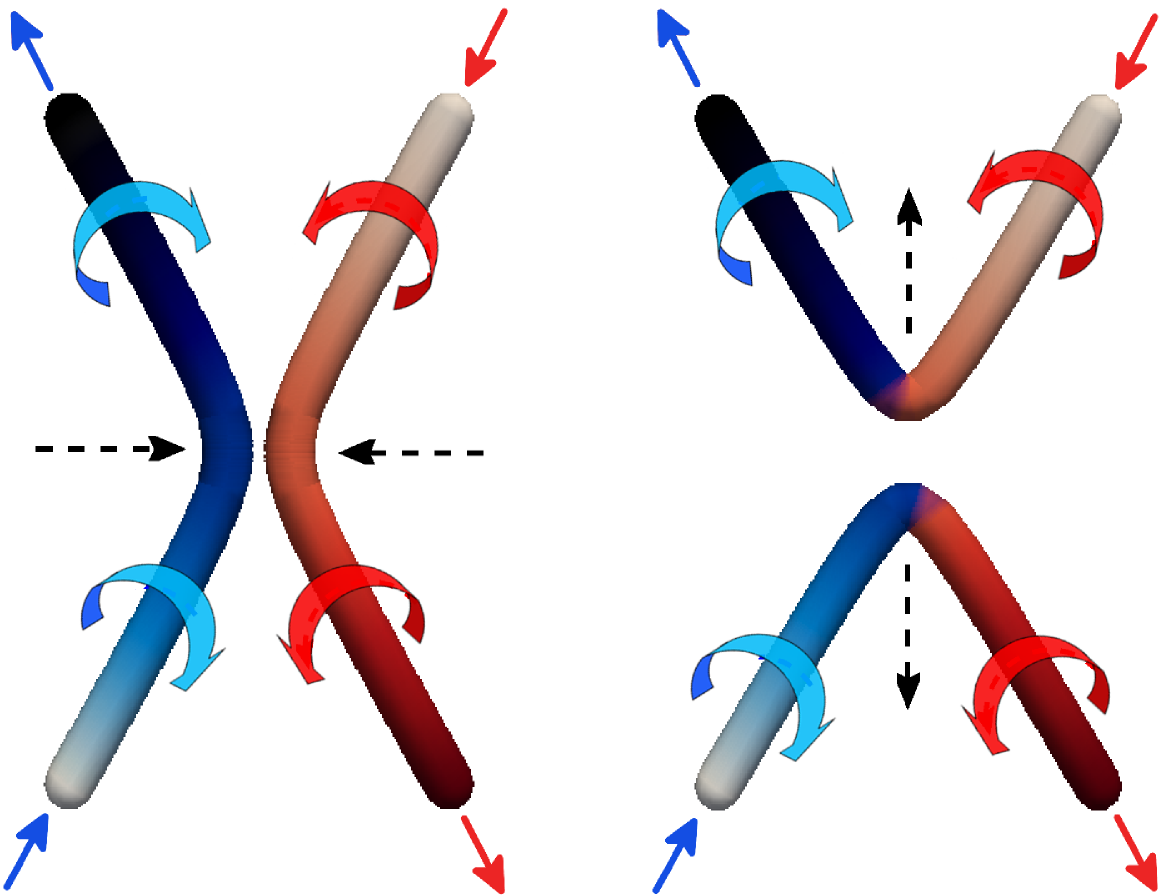}
\caption{{\bf Reconnecting vortex lines exchanging strands.}
Schematic vortex configurations before the reconnection (left)
and after (right); the vortices' shape is as
determined analytically by Nazarenko and West \cite{nazarenko-west-2003}.
Color gradient along the vortices and blue/red arrows indicate the directions
of the vorticity along the vortices and the direction of the flow
velocity around them.
Dashed black arrows indicate the vortex motion, first towards each others,
then away from each others.}
\label{fig:schematic}
\end{figure}

Reconnections of coherent filamentary structures 
(Fig.~\ref{fig:schematic}) play
a fundamental role in the dynamics of plasmas (from astrophysics 
\cite{priest-forbes-2007,che-etal-2011,cirtain-etal-2013} to confined nuclear 
fusion), nematic liquid crystals \cite{chuang-etal-1991}, polymers and 
macromolecules \cite{sumners-1995} (including DNA \cite{vazquez-sumners-2004}), optical beams \cite{dennis-etal-2010,berry-dennis-2012}, ordinary
(classical) fluids 
\cite{kida-takaoka-1994,pumir-kerr-1987,kleckner-irvine-2013} 
and quantum fluids \cite{barenghi-donnelly-vinen-2001,schwarz-1988}. 
In fluids, the coherent structures consist of concentrated vorticity, 
whose character depends on the classical or quantum nature of the fluid:
in classical fluids (air, water etc.), vorticity is a continuous field 
and the interacting structures are {\it vortex tubes}
of arbitrary core size around which the circulation of the velocity 
field is unconstrained; in quantum fluids (atomic
Bose-Einstein Condensates (BECs), and superfluid $^4$He and $^3$He), the 
structures are isolated one-dimensional {\it vortex lines}, 
corresponding to topological defects of the governing order parameter
around which the velocity's circulation is quantized \cite{onsager-1949,feynman-1955,vinen-1961,donnelly-1991}. 

The discrete nature of quantum vortices makes them ideal
for the study of vortex reconnections, which assume the form of
isolated, dramatic events, strongly localised in space and time. 
First conjectured by Feynman \cite{feynman-1955} and then numerically
predicted \cite{koplik-levine-1993}, quantum vortex reconnections  
been observed only recently, both in superfluid 
$^4$He \cite{bewley-etal-2008} (indirectly, using tracer particles) 
and in BECs \cite{serafini-etal-2017} (directly, using an 
innovative stroboscopic visualisation technique). 

Vortex reconnections are crucial in redistributing the kinetic
energy of turbulent superfluids.  In some regimes, they 
trigger a turbulent energy cascade \cite{barenghi-lvov-roche-2014}
in which vortex lines self-organise in bundles 
\cite{baggaley-laurie-barenghi-2012}, generating the same Kolmogorov 
spectrum of classical turbulence 
\cite{barenghi-lvov-roche-2014,nore-abid-brachet-1997,skrbek-sreenivasan-2012,maurer-tabeling-1998,salort-etal-2010}.  By altering the topology of 
the flow \cite{kleckner-kauffman-irvine-2016}, reconnections also
seem to redistribute its helicity 
\cite{scheeler-etal-2014,dileoni-etal-2016}, 
although the precise definition of helicity in superfluids is 
currently debated
\cite{dileoni-etal-2016,salman-2017,barenghi-etal-2018},
and the effects of reconnections 
\cite{laing-etal-2015,zuccher-ricca-2015,zuccher-ricca-2017,hanninen-etal-2016}
on its geometric ingredients (link, writhe and twist) are 
still discussed.
In the low-temperature limit, losses due to viscosity or mutual friction are 
negligible, and reconnections are the ultimate 
mechanism for the dissipation of the incompressible kinetic energy
of the superfluid via sound radiation at 
the reconnecting event \cite{leadbeater-etal-2001,zuccher-etal-2012}
followed by further sound emission by the Kelvin wave cascade 
\cite{kivotides-etal-2001,kozik-svistunov-2004, kozik-svistunov-2005} 
which follows the relaxation of the reconnection cusps.

\section*{Is there a universal route to reconnection ?}

Many authors have focused on the possibility
that there is a {\it universal route} to reconnection, which may take
the form of a vortex ring cascade \cite{kerr-2011,kursa-etal-2011}, 
a particular rule for the cusp angles \cite{dewaele-aarts-1994,tebbs-etal-2011}, 
or, more promising, a special scaling with time
of the minimum distance $\delta(t)$ between the reconnecting
vortex strands. It is on the last property that we concentrate our
attention. 

\begin{figure}[htbp]
\centering
\includegraphics[width=0.99\columnwidth]{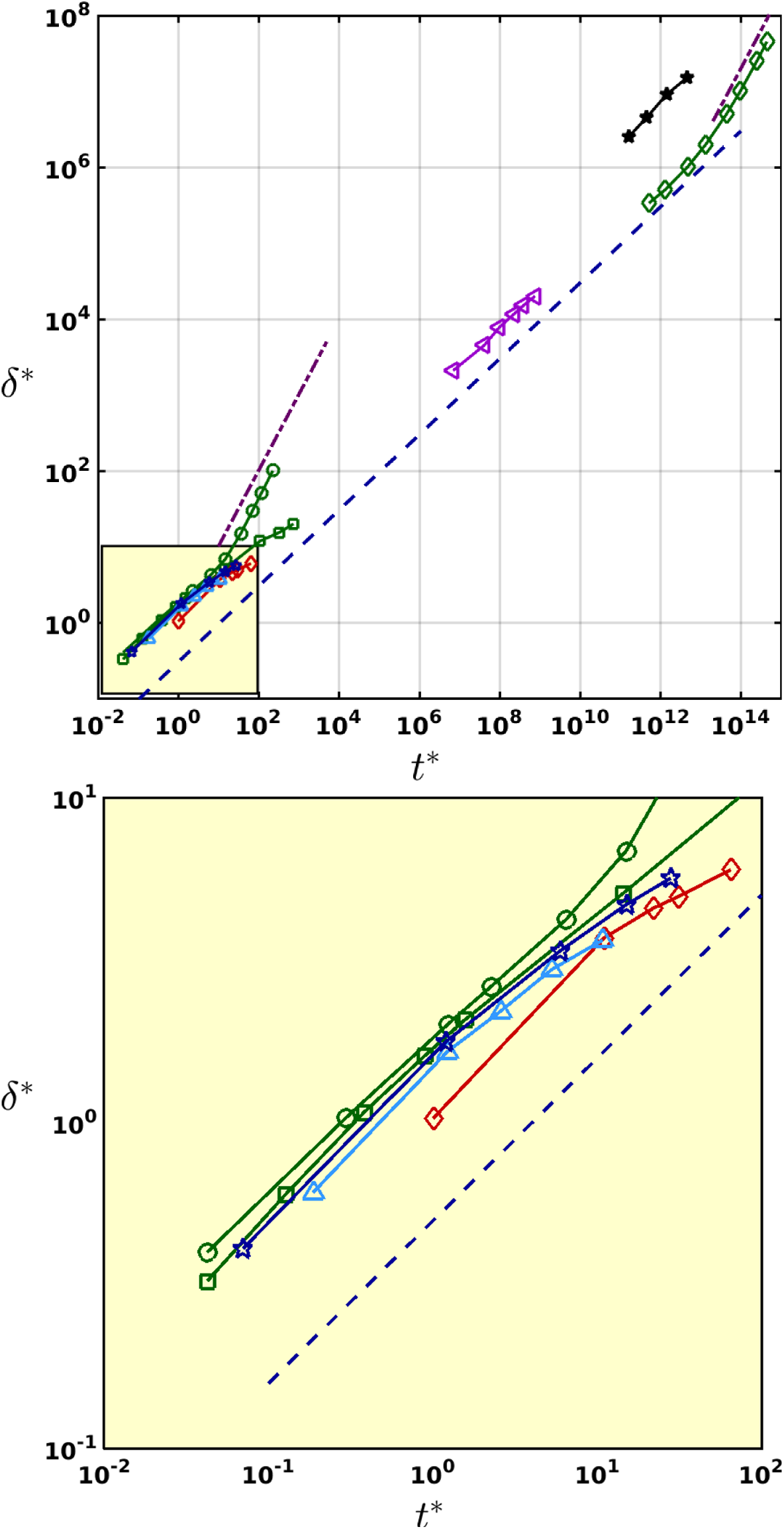}
\caption{{\bf Minimum distance between reconnecting vortices:
past and present results.} 
(top): All data reported describe the behaviour of the rescaled minimum 
distance $\delta^*$ between vortices as a function 
of the rescaled temporal distance to the reconnection event $t^*$. 
Empty (filled) symbols refer to pre (post) - reconnection dynamics. 
{\bf GP simulations}: 
\dred{$\bm{\lozenge}$} Ref.~\cite{rorai-etal-2016}; 
\dblue{{\bf \ding{73}}} Ref.~\cite{villois-etal-2017}; 
\cyan{$\bm{\triangle}$} Ref.~\cite{zuccher-etal-2012}; 
\dgreen{$\bm{\circ}$} and \dgreen{$\bm{\square}$}, present simulations, 
ring-vortex collision and orthogonal reconnection, respectively.
{\bf VFM simulations}: 
\violet{$\bm{\vartriangleleft}$} Ref.~\cite{dewaele-aarts-1994}; 
\dgreen{$\bm{\lozenge}$} present simulations, 
ring-line collision. 
{\bf Experiments:}  $\bigstar$ Ref.~\cite{fonda-etal-2019}. 
(bottom): zoom on GP simulations.
}
\label{fig:all_data}
\end{figure}

Several studies have observed a symmetrical
pre/post reconnection scaling of $\delta(t)$
\cite{dewaele-aarts-1994,nazarenko-west-2003,paoletti-etal-2010,dossantos-2016,villois-etal-2017}; 
others have suggested an asymmetrical scaling possibly ascribed to 
acoustic energy losses 
\cite{zuccher-etal-2012,allen-etal-2014,rorai-etal-2016},
similar to the asymmetry observed in classical Navier-Stokes 
fluids \cite{hussain-duraisamy-2011}. 
In Fig.~\ref{fig:all_data} (top) and (bottom) we present a comprehensive
summary of the scaling of $\delta(t)$, combining previous numerical and experimental 
results with data computed in the present study; this spans an impressive eight orders of magnitude.
  
The aim of this paper is to reveal that there are 
{\it two} distinct fundamental scaling regimes for
$\delta(t)$. In addition to the known 
\cite{dewaele-aarts-1994,nazarenko-west-2003,paoletti-etal-2010,dossantos-2016,villois-etal-2017,fonda-etal-2019,fonda-etal-2014,lipniacki-2000}
$\delta \sim t^{1/2}$ scaling,
we predict and observe a new linear scaling $\delta(t) \sim t $. 
We show how the two scalings arise from rigorous
dimensional arguments, then demonstrate them
in numerical simulations of vortex reconnections


\section*{Dimensional analysis}

We conjecture that, in the system under consideration (superfluid helium,
atomic BECs), $\delta$ depends only upon the following physical 
variables: the time $t$ from the reconnection, 
the quantum of circulation $\kappa$ of the superfluid, a characteristic 
lengthscale $\ell$ associated to the geometry of the vortex configuration,
the fluid's density $\rho$, and 
the density gradient $\nabla\rho$.  
We hence postulate the following functional form
\begin{equation}
\displaystyle 
f (\delta, t, \kappa, \ell, \rho, \nabla\rho) = 0   \; .\label{eq:dim} 
\end{equation}
Following the standard procedure of the Buckingham $\pi$-theorem  \cite{buckingham-1914} (see SI Appendix SI.1
for details),
we derive the following scalings:
\begin{eqnarray}
\displaystyle
\delta(t) &=& (C_1\kappa)^{1/2}t^{1/2}~~~~  \textrm{\it interaction regime}\label{eq:dim.1/2},\\[2mm]
\delta(t) &=& C_2 \left ( \frac{\kappa}{\ell} \right ) t ~~~~~~~~ \textrm{\it driven regime}\label{eq:dim.ell}, \;\;\;\;\\[2mm]
\delta(t) &=& C_3 \left (\kappa\frac{\nabla\rho}{\rho} \right ) t ~~~~ \textrm{\it driven regime}\label{eq:dim.rho}, 
\end{eqnarray}
where $C_1$, $C_2$ and $C_3$ are dimensionless constants. 
Physically, the $\delta \sim t^{1/2}$ scaling of Eq. (\ref{eq:dim.1/2}) 
identifies 
the quantum of circulation $\kappa$ as the only relevant parameter 
driving the reconnection
dynamics \cite{bewley-etal-2008}; 
this scaling
corresponds to a vortex dynamics driven by the mutual interaction between vortex strands, as illustrated more in detail in
section A.

Equations (\ref{eq:dim.ell}, \ref{eq:dim.rho}), on the other hand, introduce the new $\delta \sim t$ scaling.  This scaling suggests the presence of a characteristic velocity which drives the approach/separation of the vortex lines. 
Indeed, we can offer some physical examples of these velocities.  If
$\ell$ is the radius of a vortex ring, then
$v_{\ell}=C_2(\kappa/\ell)$ is, 
to a first approximation, the self-induced velocity of the ring.  Alternatively, if
$\ell$ is equal to the distance of a vortex to a sharp boundary in an otherwise
homogeneous BEC (such as arises for BEC confined by box traps), then $v_{\ell}$ is the self-induced vortex
velocity arising from the presence of an image vortex.  Finally,
if the BEC density is smoothly-varying (such as arises for BECs confined by harmonic traps) 
then $v_{_{\nabla\!\rho}}=C_3(\kappa\nabla\rho/\rho)$ is 
precisely the individual velocity of a vortex induced by the 
density gradients, $C_3$ depending on the trap's geometry 
\cite{jackson-etal-1999,svidzinsky-fetter-2000}. 
In the next section we will see how these scalings, and the crossover, emerge in typical scenarios through numerical simulations.


\section*{Numerical simulations}

There are two established models of quantum vortex dynamics:
the Gross-Pitaevskii (GP) model and the Vortex Filament (VF) method.
The former describes a weakly-interacting BEC in the zero-temperature 
limit \cite{pitaevskii-stringari-2003}, the latter is based on the
classical Biot-Savart law describing the velocity field of a given 
vorticity distribution, which in our case is concentrated on space curves
\cite{schwarz-1985,hanninen-baggaley-2014}.


The main difference between GP and VF models is the probed 
lengthscales of the flow. The GP equation is a {\it microscopic}, 
compressible model,
capable of describing density fluctuations
and lengthscales smaller than
the vortex core $a_0$
(defined as the diameter of the cylindrical tube around the superfluid vortex-line
where the density is within $75\%$ of the bulk density).
In the GP model, vortices are identified 
as topological phase defects of the condensate wavefunction $\Psi$, and 
reconnections are solutions of the GP equation itself.
On the other hand, the VF method is a {\it mesoscopic} incompressible model, 
probing the features of the flow at lengthscales much larger
than the vortex core, typically 
$10^4\,a_0$ or $10^5\,a_0$, neglecting 
any density perturbation created by moving vortices and the density
depletions represented by the vortex cores themselves.
In the VF model, vortex lines
are discretised employing a set of Lagrangian points whose dynamics is 
governed by the classical Biot-Savart law, and vortex reconnections are
performed by an
{\it ad hoc} `cut-and-paste' algorithm \cite{schwarz-1985,baggaley-2012b}. 

In the present study, we employ both GP and VF models to investigate 
the scaling with time of the minimum distance $\delta$ between 
reconnecting vortices. Technical details of these methods are described in SI Appendix SI.6 and SI.7.
Distinctive of our simulations is the larger initial distance $\delta_0$ compared to past numerical studies (5 to 20 times larger in GP simulations,
100 to 2000 larger in VF ones). We also extend the use of the GP model
to inhomogeneous, confined BECs where  vortex reconnections 
can now be investigate experimentally with 
unprecedented resolution \cite{serafini-etal-2017}.  

\begin{figure*}[htbp]
\centering
\includegraphics[width=0.99\textwidth]{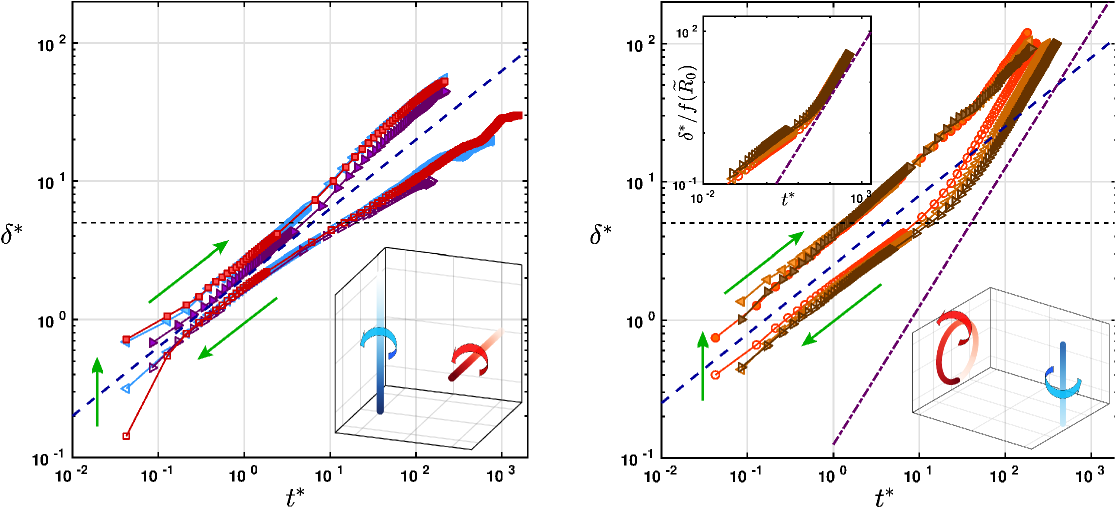}
\caption{{\bf GP simulations: homogeneous unbounded BECs.}
Evolution of the rescaled minimum distance $\delta^*$ between reconnecting vortices as a function of the rescaled temporal distance to reconnection $t^*$. 
Empty (filled) symbols correspond to pre (post) reconnection dynamics. {\bf Left}: orthogonal vortices reconnection with rescaled initial distance $\delta^*_0$ equal to
$10$ (\violet{$\bm{\vartriangleright}$}), $20$ (\cyan{$\bm{\vartriangleleft}$}) and $30$ (\dred{$\bm{\square}$}). 
{\bf Right}: ring-line reconnection for constant initial distance $\delta^*_0=100$ and vortex ring radii $R^*_0$ equal to $5$ 
(\orange{$\bm{\circ}$}), $7.5$ (\dyellow{$\bm{\vartriangleleft}$}) and $10$ (\brown{$\bm{\vartriangleright}$}). {\bf Inset}: pre-reconnection dynamics only, the distance is rescaled with $f(\widetilde{R}_0)$.
In both sub figures: the horizontal dashed black line indicates the width of the vortex core ($\approx 5\,\xi$), the blue-dashed line shows the $t^{*^{1/2}}$ scaling and the bottom insets show the initial
vortex configuration. Color gradient on vortices indicates direction of the superfluid vorticity (from light to dark). Dot-dashed violet line in the
right panel indicates the $t^{\ast}$ scaling.  Green arrows indicate the direction of time.} 
\label{fig:GP_homog}
\end{figure*}

\subsection{Homogeneous unbounded systems}

To make progress in the understanding of vortex reconnections in homogenous 
quantum fluids, we identify two limiting initial vortex configurations
which generate the two fundamental types of reconnections. The first
configuration consists of
two initially straight and orthogonal vortices, corresponding to the limit
where the curvatures $K_1$ and $K_2$ of the two vortices are small and
comparable ({\it i.e.} $K_1 \sim K_2$ and $K_1, K_2 \ll 1$);
the second configuration is
a vortex ring interacting with an isolated vortex line,
which is the limiting case of
two vortices of significantly different
curvatures ($K_1 \ll K_2$ or $K_1 \gg K_2$). The
third limiting case of large and comparable curvatures ($K_1 \sim K_2$ and $K_1, K_2 \gg 1$), 
{\it i.e.} the collision of small vortex rings, is neglected in the present study as it refers to an extremely unlikely event, due to the small cross section.

The orthogonal reconnection configuration, and the corresponding results for $\delta(t)$, 
are shown in Fig.~\ref{fig:GP_homog} (left) and reported in movies S1-S4.
Previous GP simulations of this geometry used
initial distances $\delta_0 \lesssim 6\,\xi$, where $\xi=\hbar/\sqrt{2 m g n}$ 
is the healing length of the system ($a_0\approx 4\,\xi$ to $5\,\xi$), 
and $m$, $g$ and $n$ are the boson mass,
the repulsive strength of boson interaction and the bulk
density of bosons respectively.
Here we extend the investigations to initial distances 
$\delta_0 \approx 30\,\xi$. 
Introducing the rescaled distance $\delta^*=\delta/\xi$ and 
time $t^*=|t-t_r|/\tau$ (where $t_r$ is the 
reconnection instant and $\tau=\xi/c$, $c=\sqrt{g n/m}$ being 
the speed of sound in a homogeneous BEC), 
we observe that for $\delta^*\lesssim 2.5$ 
(when the two vortex lines are so close to each other that the
condensate's density in the region between them is significantly less
than the bulk density)
a symmetrical $t^{\ast^{1/2}}$ scaling emerges clearly for both pre- and post-reconnection dynamics.  This is consistent with the most recent
GP simulations \cite{villois-etal-2017} and inconsistent with other numerical GP studies \cite{zuccher-etal-2012,rorai-etal-2016},
adding further evidence to a $t^{\ast^{1/2}}$ symmetrical scaling at small distances for orthogonal reconnections.

\begin{table}[htbp]
\begin{center}
\begin{tabular}{ c | c | c | c |}
             & $\delta^*< 2.5$ & $2.5 < \delta^* < 10$ & $\delta^*> 10$  \\
  \hline
  \hline           
  $\alpha^-$ &      0.48       &         0.40          &      0.43       \\
  $\alpha^+$ &      0.49       &         0.60          &      0.50       \\
\end{tabular}
\end{center}
\caption{{\bf GP simulations: homogeneous unbounded BECs, orthogonal reconnection}.
\textmd{\footnotesize Scaling exponents $\alpha^-$ ($\alpha^+$) for power law behaviour $\delta^* \sim t^{*^{\alpha}}$ for pre (post) reconnection
dynamics for initial separation $\delta_0^*=20$.}}
\label{tab:fits}
\end{table}

\begin{figure}[htbp]
\includegraphics[width=0.99\columnwidth]{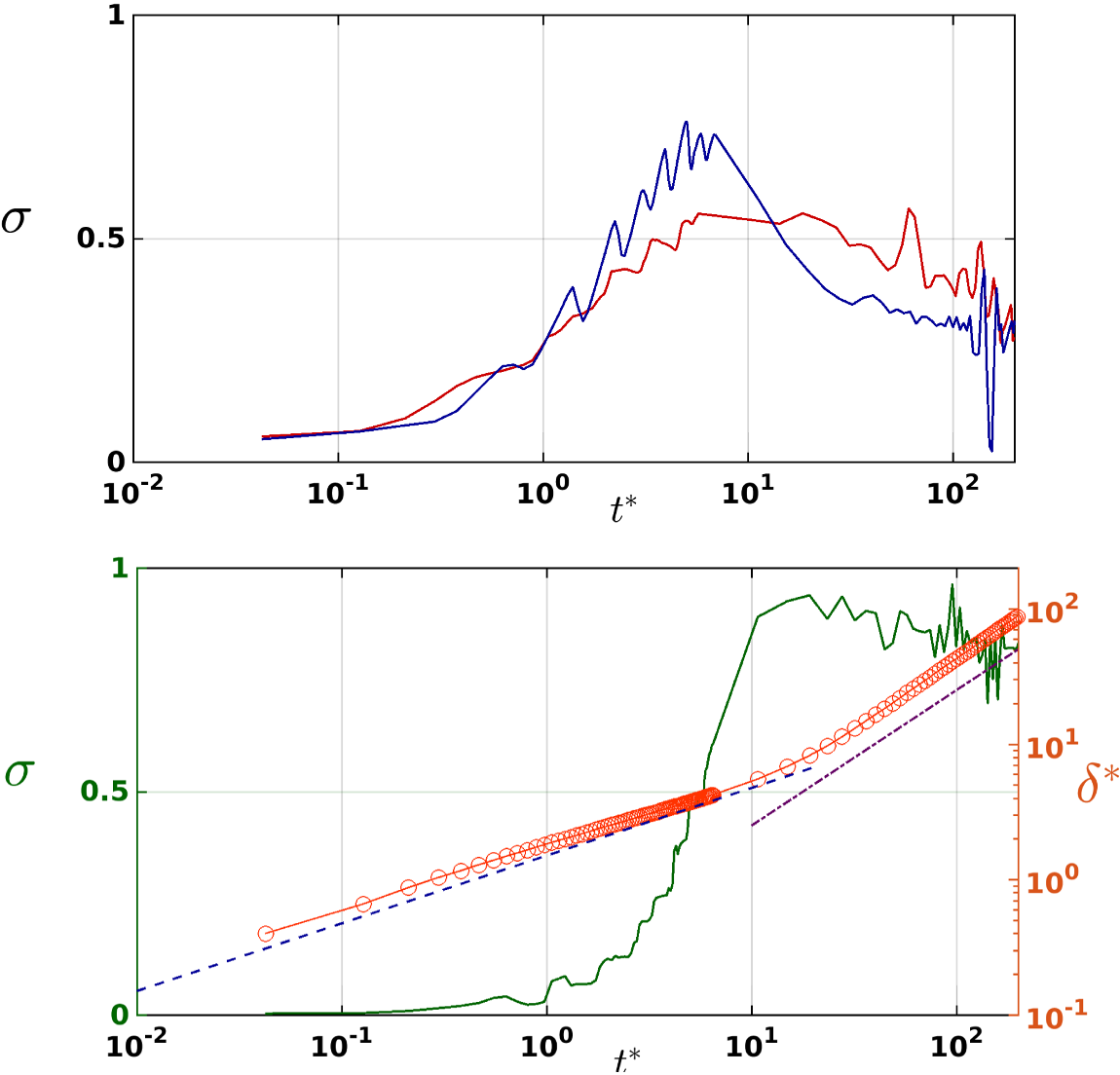}
\caption{{\bf Curvature contribution.} 
(Top): Temporal evolution of the ratio $\sigma$ for the orthogonal reconnection with initial separation distance 
$\delta_0^*~=~20$; red (blue) symbols correspond to pre (post) reconnection dynamics.
(Bottom): Temporal evolution of $\sigma$ (green solid line) and rescaled minimum distance $\delta^*$ (orange circles) 
for the ring-line pre-reconnection, with $R^*_0~=~5$. The blue-dashed line shows the $t^{*^{1/2}}$ scaling, while dot-dashed violet line
indicates the $t^{\ast}$ scaling.}
\label{fig:GP_curv_contr}
\end{figure}

To map quantitatively
the emergence of the
$t^{\ast^{1/2}}$ behaviour in distinct intervals of $\delta$, in Table \ref{tab:fits}
we report the scaling exponents $\alpha$ of the power-law fits $\delta^*\sim t^{*^{\alpha}}$ for the intermediate initial distance $\delta_0^*=20$.
From the table it clearly emerges that
the $t^{\ast^{1/2}}$ scaling also holds in the post-reconnection dynamics at large distances, while in the intermediate region $2.5 < \delta^* < 10$ (both pre- and post
reconnection) and at large distances 
during the approach, the scalings deviate from $t^{\ast^{1/2}}$. 
In order to investigate these deviations, 
we calculate the
velocity 
contribution of the local 
vortex curvature, $v_{\gamma}$, to the approach/separation velocity $d \delta^*/d t^*$  for the intermediate initial distance $\delta_0^*=20$,
displaying the corresponding results in Fig.~\ref{fig:GP_curv_contr} (top)
(see SI Appendix SI.2 for the calculation of $v_{\gamma}$).
The figure clearly shows that
the observed deviations from the
$t^{\ast^{1/2}}$ scaling depend on the relative curvature contribution $\sigma~=~v_{\gamma}/(d \delta^*/d t^*)$:
the larger $\sigma$, the more prominent the deviations from the $t^{\ast^{1/2}}$ behaviour. This implies that the $t^{\ast^{1/2}}$ scaling
observed at large distances in the separation dynamics originates from an interaction-dominated motion of the reconnecting strands. If 
the dynamics is governed by the mutual interaction of the two vortices, in fact,  
$d \delta^*/d t^* \propto \kappa/(4\pi\delta^*)$ \cite{dewaele-aarts-1994},
leading straightforwardly to the scaling $\delta^*\sim t^{\ast^{1/2}}$, derived in Eq. (\ref{eq:dim.1/2}). 
This argument 
corroborates
the experimentally observed $t^{\ast^{1/2}}$ 
scaling \cite{paoletti-etal-2010,fonda-etal-2019}.
Concluding our
study of orthogonal reconnections, we note that vortex 
lines move faster after the
reconnection than before it, as pointed out in past studies \cite{zuccher-etal-2012,allen-etal-2014,rorai-etal-2016,villois-etal-2017}, and
that the post-reconnection curves show a slight sensitivity to the initial distance $\delta_0^*$.

The second homogeneous system which we consider is a vortex ring reconnecting with
an isolated initially straight vortex line (see movies S5-S8).
This scenario is notably relevant in superfluid helium turbulence at low temperatures, where the low density
of thermal excitations is not able to quickly damp out small 
vortex structures.
In fact, the methods used to generate turbulence at low 
temperatures involve
either injecting vortex rings 
\cite{walmsley-golov-2008,walmsley-etal-2014,zmeev-etal-2015,
walmsley-etal-2017} or vibrating small
structures ({\it e.g.} spheres, wires and grids) which trigger a great number of ring - line collisions
\cite{hanninen-etal-2007,goto-etal-2008,nakatsuji-etal-2014}. Similarly, also inhomogeneous quantum turbulence
\cite{nemirovskii-2010,barenghi-samuels-2002} and boundary layer turbulence \cite{stagg-parker-barenghi-2017} 
display conspicuous vortex loop - vortex line collisions.

Figure~\ref{fig:GP_homog} (right) illustrates this vortex set-up and presents the behaviour of $\delta(t)$.  
We vary the initial radius of the ring ($R^*_0=R_0/\xi=5,7.5$ and $10$), while keeping constant its
initial distance $\delta^*_0=100$ to the line. 
We first focus on the pre-reconnection evolution of $\delta(t)$.  This
clearly reveals the two distinct scalings predicted by the
dimensional analysis:
\begin{eqnarray}
\displaystyle
\delta^*\!(t^*) & \sim &  a_{_{1/2}} t^{\ast^{1/2}} \;\;\; \text{for} \;\;\; t^* \lesssim 5 \; ,\label{eq:scal.0.5}\\\nonumber\\
\delta^*\!(t^*) & \sim &  a_{_1} t^{\ast}             \,\;\;\; \;\;  \;\; \text{for} \;\;\; t^* \gtrsim 20 \; ,\label{eq:scal.1}
\end{eqnarray} 
\noindent
where $a_{_{1/2}}$ and $a_{_{1}}$ are
constant prefactors corresponding, respectively, to $(C_1\kappa)^{1/2}$ in Eq.~(\ref{eq:dim.1/2}) and $v_\ell$ or $v_{_{\nabla\!\rho}}$ 
in Eqs.~(\ref{eq:dim.ell}) or (\ref{eq:dim.rho}).
To our knowledge, the linear scaling has 
not been reported in previous studies. We also note that the crossover between these two regimes occurs at a distance of $\delta_c \sim 3-4 \, \xi$; 
we will revisit the importance of this scale later.

The linear scaling
implies that $d \delta^*/d t^*$ is constant: 
at large distances, the relative velocity between the two points at 
minimum distance, $\mathbf{x}_{\rm ring}$ (on the vortex ring) 
and $\mathbf{x}_{\rm line}$ (on the vortex line),
projected on the separation vector 
$\bm{\delta}=\mathbf{x}_{\rm ring}-\mathbf{x}_{\rm line}$,
is constant with respect to time. We argue that, at large
separation distances
$d \delta^*/d t^*$ is approximately equal to the initial speed
of the vortex ring
\cite{roberts-grant-1971}:
\begin{equation}
\displaystyle \frac{d \delta^*}{d t^*} = v_{\rm ring}^* = \frac{\kappa}{4\pi \xi c R_0^*}\bigg [ \ln (8 R_0^*) - 0.615  \bigg ]\,\,\, \label{eq:delta.der}. 
\end{equation} 
The self-induced velocity of the vortex ring $v_{\rm ring}$ thus plays 
the role of the characteristic velocity $v_\ell$ 
in
Eq.~(\ref{eq:dim.ell}).
We make the notation compact and rewrite Eq.~(\ref{eq:delta.der}) as
$v_{\rm ring}^*~=~Cf(\widetilde{R}_0)$, where $C~=~\kappa/(4\pi c\overline{R}_0)$, 
$f(\widetilde{R}_0)~=~[\ln(\widetilde{R}_0) + 3.48] /\widetilde{R}_0$, 
$\overline{R}_0=7.5\,\xi$ is
the average radius of the three simulations, 
and $\widetilde{R}_0=R_0/\bar{R}_0$.  We then arrive at the result that Eq.~(\ref{eq:scal.1}) 
takes the form 
$\displaystyle \delta^*\!(t^*) \sim C f(\widetilde{R}_0)\, t^*$. 
This is confirmed in Fig.~\ref{fig:GP_homog} (right, inset): when $\delta^*$ is rescaled as 
$\delta^*/f(\widetilde{R}_0)$ 
the curves collapse onto a single curve in the $\delta \sim t$ regime.
The clear $t^{\ast^{1/2}}$ scaling for $\delta^*\lesssim \delta_c^*$ is consistent with the interpretation put forward for the orthogonal reconnection,
as at such small distances, the approach/separation is likely to be governed by the mutual interaction of the two vortices
given that $\delta$ is smaller than the ring's radius of curvature.

We hence suggest that these two scalings correspond to a crossover from 
a dynamics predominantly driven by mutual vortex interaction
($\delta^\ast \sim t^{\ast^{1/2}}$ scaling) 
to the self-driven (curvature-driven) motion of the ring ($\delta^\ast \sim t^\ast$ scaling).
To check this conjecture, we again refer to the contribution of the local vortex curvature:
Fig.~\ref{fig:GP_curv_contr} (bottom) shows the relative curvature contribution for the ring-line pre-reconnection dynamics.
We see that the contribution from the local vortex curvature to the approach velocity
drops dramatically for $t^*\lesssim 5$,
corresponding exactly to the onset of the $t^{\ast^{1/2}}$ scaling, supporting this picture. 

The crossover between the two scalings, however,
is less apparent in the post-reconnection dynamics, and for two main reasons.  Firstly, both vortices become perturbed by propagating Kelvin-waves;
secondly, the travelling velocity of the perturbed vortex ring is not constant 
\cite{arms-hama-1965,dhanak-debernardinis-1981,barenghi-hanninen-tsubota-2006}. 
These Kelvin waves 
generate sound waves \cite{kopiev-Chernyshev-1997,parker-etal-2004} 
dissipating the incompressible kinetic energy, leading to a decrease of 
the length of the vortex ring and a damping of the oscillations' amplitude.
When these oscillations die out
({\it e.g.} in the simulation with $R^*_0=5$, see Fig. \ref{fig:GP_homog} (right)),
and the vortex ring regains its circular shape travelling at 
constant velocity away from the vortex line, we
recover the expected $\delta^{\ast} \sim t^{\ast}$ scaling. 
This wobbling dynamics and the recovery of the linear scaling are addressed in more detail in 
SI Appendix SI.3.

The same qualitative behaviour for the orthogonal vortices and ring-line scenario is recovered in 
VF simulations (see SI Appendix SI.4, Fig.~S2).  In the latter scenario, 
the crossover from $t^{\ast^{1/2}}$ to $t^{\ast}$ scaling occurs at much larger lengthscales than in the GP simulations,
given the range of scales involved ($\approx 10^5 a_0 - 10^8 a_0$).  
However, as for GP simulations, the distance $\delta_{\rm c}$ at which the crossover takes place is determined by
the balance between interaction-dominated motion and curvature-driven dynamics, {\it i.e.} by the comparison
between $\delta(t)$ and the radius of curvature $R_c(t)$ of the vortex ring. 


\begin{figure}[ht]
\centering
\includegraphics[width=0.9\columnwidth]{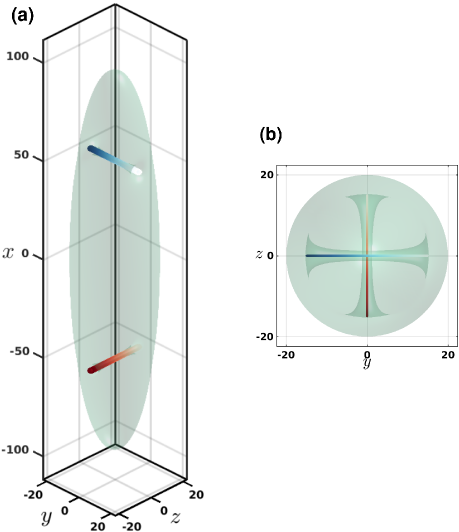}
\caption{{\bf GP simulations: harmonically trapped BECs, initial conditions}.
{\bf (a), (b)}: Lateral and top view of initial vortex configuration. 
Light green surfaces are isosurfaces of condensate density at $5\%$ of trap-center density. Color gradient on vortices indicates the direction of the superfluid vorticity (from light to dark).
Unit of length is the healing length $\xi_c$ evaluated in the center of the trap.} 
\label{fig:harm.traps.ic}
\end{figure}

\subsection{Trapped systems}

\begin{figure*}[ht]
\centering
\includegraphics[width=0.9\textwidth]{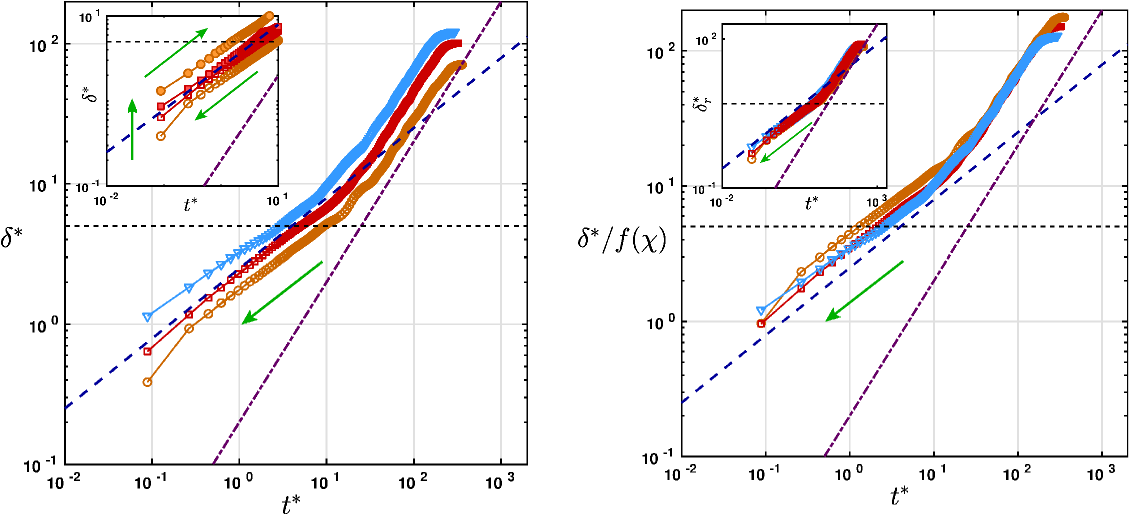}
\caption{{\bf GP simulations: harmonically trapped BECs}.
Evolution of the minimum distance $\delta^*$ between reconnecting vortices as a function of the temporal distance to reconnection $t^*$. 
Open (filled) symbols correspond to pre (post) reconnection dynamics. {\bf Left}: pre-reconnection scaling of $\delta^*$ for initially 
imprinted orthogonal vortices with corresponding orbit parameter $\chi=0.35$ (\dyellow{$\bm{\circ}$}), $\chi=0.5$ (\dred{$\bm{\square}$}) and $\chi=0.6$ (\cyan{$\bm{\bigtriangledown}$}). 
Inset: short-time pre-(open symbols) and post-(filled symbols) reconnection scaling of $\delta^*$ for $\chi=0.35$ (\dyellow{$\bm{\circ}$}), and $\chi=0.5$ (\dred{$\bm{\square}$}). 
{\bf Right}: temporal evolution of the minimum distance $\delta^*$ rescaled with $f(\chi)$. Symbols as in left panel.
Inset: short-time pre-reconnection scaling of rescaled minimum distance $\delta^*_r=\delta/\xi_r$.   
In both subfigures, the dashed blue and dot-dashed violet lines show the $t^{*^{1/2}}$ and $t^{\ast}$ 
scalings, respectively. The horizontal dashed line indicates the width of the vortex core at the center of the trap ($\approx 5\,\xi$). Green arrows indicate the direction of time.} 
\label{fig:harm.traps}
\end{figure*}

Since it is now experimentally
possible to visualise individual quantum reconnections
in trapped atomic BECs \cite{serafini-etal-2015,serafini-etal-2017}, we
test the above results under such realistic
experimental set-ups.
We consider two classes of traps: 
the widely-employed {\it harmonic traps} \cite{dalfovo-etal-1999} 
(Figs.~\ref{fig:harm.traps.ic} and \ref{fig:harm.traps} and movies S9 and S10) 
and the recently-designed {\it box-traps} 
\cite{gaunt-etal-2013,navon-etal-2016} (see SI Appendix SI.5 and movies S11 and S12).
We employ GP simulations throughout this analysis (the VF model is not suitable for inhomogeneous systems).

In harmonic traps the condensate is inhomogeneous (the density
is larger near the centre) and individual motion of the vortices 
(responsible for the linear scaling) is determined
by their curvature, density gradients and, possibly, vortex images
\cite{svidzinsky-fetter-2000,fetter-svidzinsky-2001,jackson-etal-1999,fetter-2009}.
In box-traps the condensate's density is constant (with the
exception of a thin layer of width of the order of the healing length near 
the boundary), and
the individual vortex motion is believed to be driven by image vortices with respect to 
the boundaries, according to two-dimensional studies \cite{mason-berloff-fetter-2006}.
We exploit these self-driven vortex motions to analyse reconnections 
starting from initial distances significantly larger 
than previous numerical simulations (up to 20 times larger).

Consider first the harmonic trap case; the initial configuration is shown in Fig.~\ref{fig:harm.traps.ic}.
The condensate is taken to be cigar-shaped, with the long axis along $x$ (the trapping
frequency $\omega_x$ along $x$ is smaller than those in the transverse directions,
$\omega_y=\omega_z=\omega_{\perp}$).
In this geometry, a single straight vortex line imprinted off center 
on a radial plane is known to orbit around the center of the condensate 
\cite{anderson-etal-cornell-2000,freilich-etal-2010} 
along an elliptical orbit perpendicular to itself.  The vortex follows
a trajectory of constant energy \cite{svidzinsky-fetter-2000} which
is uniquely determined by the orbit parameter $\chi=x_0/R_x=r_0/R_\perp$, 
where $x_0$ and $r_0$ are the axial and radial semi-axes of the ellipse, 
and $R_x$ and $R_\perp$ are the axial and radial Thomas-Fermi radii 
respectively.  The period $T$ of this orbit decreases with increasing 
$\chi$ \cite{serafini-etal-2015,svidzinsky-fetter-2000,lundh-ao-2000,sheehy-radzihovsky-2004,fetter-2009},
$\displaystyle T=\frac{8 \pi \left ( 1 - \chi^2 \right ) \mu}{3\hbar \omega_\perp \omega_x \ln (R_\perp/\xi_c)}$, where $\xi_c$ is the 
healing length at the center of the trap. Hence, outer vortices (with larger values of $\chi$) move faster.


If two orthogonal vortices are imprinted on radial planes, 
intersecting the (long) $x$ axis at opposite positions $\pm x_0$,
distinct vortex interactions can occur (vortex rebounds, vortex
reconnections, double reconnections, ejections) 
depending on the value of the orbit parameter 
$\chi$ \cite{serafini-etal-2017}. 
Results presented here refer to
three different values of $\chi$, all engendering vortex reconnections:
$\chi=0.35,\, 0.5,\, 0.6$. The pre-reconnection evolution of $\delta^*=\delta/\xi_c$ is reported in 
Fig.~\ref{fig:harm.traps} (left).   As for the ring-line reconnection, we observe a cross-over from
$t^*{^{1/2}}$ to $t^{\ast}$ scaling.  This occurs for all values of $\chi$.  Moreover, the $t^{*^{1/2}}$ scaling again occurs in the region $\delta^*\lesssim 5$,  
suggesting that this feature is indeed universal
for vortex reconnections in BECs. 

If we rescale the minimum distance $\delta$ with the 
healing length $\xi_r$ evaluated at the reconnection point $\mathbf{x}_r$,
$\displaystyle \xi_r=\frac{\hbar}{\sqrt{2 g m n_{_{\rm TF}}(\mathbf{x}_r)}}$ ($n_{_{\rm TF}}(\mathbf{x}_r)$ being the condensate particle density according to the Thomas-Fermi approximation), the curves corresponding to 
distinct values of $\chi$ overlap for $\delta_r^*=\delta/\xi_r \lesssim 5$
(see the inset of the right panel of Fig.~\ref{fig:harm.traps}).
This result implies that $\xi_r$ (hence the radius of the 
vortex core) is the correct lengthscale which characterises
the approach dynamics when vortex cores start merging. 
Furthermore, the dependence of $\xi_r$ on $m n_{_{\rm TF}}(\mathbf{x}_r)$ 
indicates that mass density $\rho= m n$ itself plays a significant role in 
determining the minimum distance $\delta$  - this is exactly why we 
included $\rho$ in the set of physical variables when applying
Buckingham's $\pi$-Theorem.

Another similarity between reconnections in harmonic traps and
other geometries is the faster post-reconnection dynamics, 
as seen in the inset of Fig.~\ref{fig:harm.traps} (left). This velocity difference
between approach and separation (related to an increase of the local vortex 
curvature in the reconnection process and to an emission of 
acoustic energy) seems
a universal feature of quantum vortex reconnections 
\cite{villois-etal-2017}
and also observed in simulations of reconnecting
classical vortex tubes \cite{hussain-duraisamy-2011}.    

Figure~\ref{fig:harm.traps} (left) shows that 
$\displaystyle \, \frac{d \delta^*}{d t^*}$ is constant
for $t^*\gtrsim 20$ before the reconnection, increasing with
increasing values of the orbital parameter $\chi$ 
(this is not surprising since isolated vortices move faster on 
outer orbits). 
It seems reasonable to assume that
$\displaystyle \, \frac{d \delta^*}{d t^*} = C f(\chi)$, 
where $\displaystyle f(\chi) = \frac{\chi}{1-\chi^2}$ and $C$ 
is a constant which depends on the trap's geometry. Indeed, 
the magnitude of the vortex velocity induced by both density gradients 
\cite{svidzinsky-fetter-2000,svidzinsky-fetter-2000PRA,fetter-svidzinsky-2001}
and vortex curvature (assuming, for simplicity, that the shape of the
vortex is an arc of a circle) are proportional to $f(\chi)$. 
As a consequence, we expect that
$\delta^*(t^*)\sim C f(\chi)\, t^*$ for $t^*\gtrsim 20$.
This conjecture is confirmed in Fig.~\ref{fig:harm.traps} (right): 
when plotted as $\delta^*/f(\chi)$, the curves for different $\chi$ collapse onto a universal curve in this region.

We stress that the observed linear scaling at large distances is a novel result. However,
although we have numerically
identified the dependence of $d \delta^*/d t^*$ on $\chi$ at large
distances, we still lack a simple physical justification of this result.

In harmonic traps, 
the predominant effect driving the approach of the vortices at large
distances is hence the individual vortex motion driven by curvature and  
density gradients (the role of vortex images still remains unclear \cite{fetter-2009} in this trap geometry), 
independent of the presence of the other vortex. 
The scaling crossover in harmonic traps
is thus governed by the balance between the interaction of the reconnecting strands and the driving of the individual vortices,
as it occurs for the ring - line reconnection in homogeneous BECs described previously. 

The nature of this scaling crossover is confirmed by the investigation of vortex reconnections 
in box-trapped BECs, outlined in SI Appendix SI.5 (see, in particular, Fig.~S4). In these
trapped systems, the motion of individual vortices is found to be driven by vortex images, leading to a linear scaling
at large distances. At small distance we again recover the $\delta \sim  t^{\ast^{1/2}}$ scaling. 
The results obtained in all the trapped BECs investigated in this work, hence, always show a $\delta \sim  t^{\ast^{1/2}}$ to $\delta \sim  t^{\ast}$ scaling
crossover which, we stress, has not been observed in past studies. In addition, the always observed small scale $\delta \sim  t^{\ast^{1/2}}$ 
behaviour supports the argument for
the existence of a universal scaling law at length scales
close to the reconnection point.


\begin{figure}[htbp]
\centering
\includegraphics[width=0.94\columnwidth]{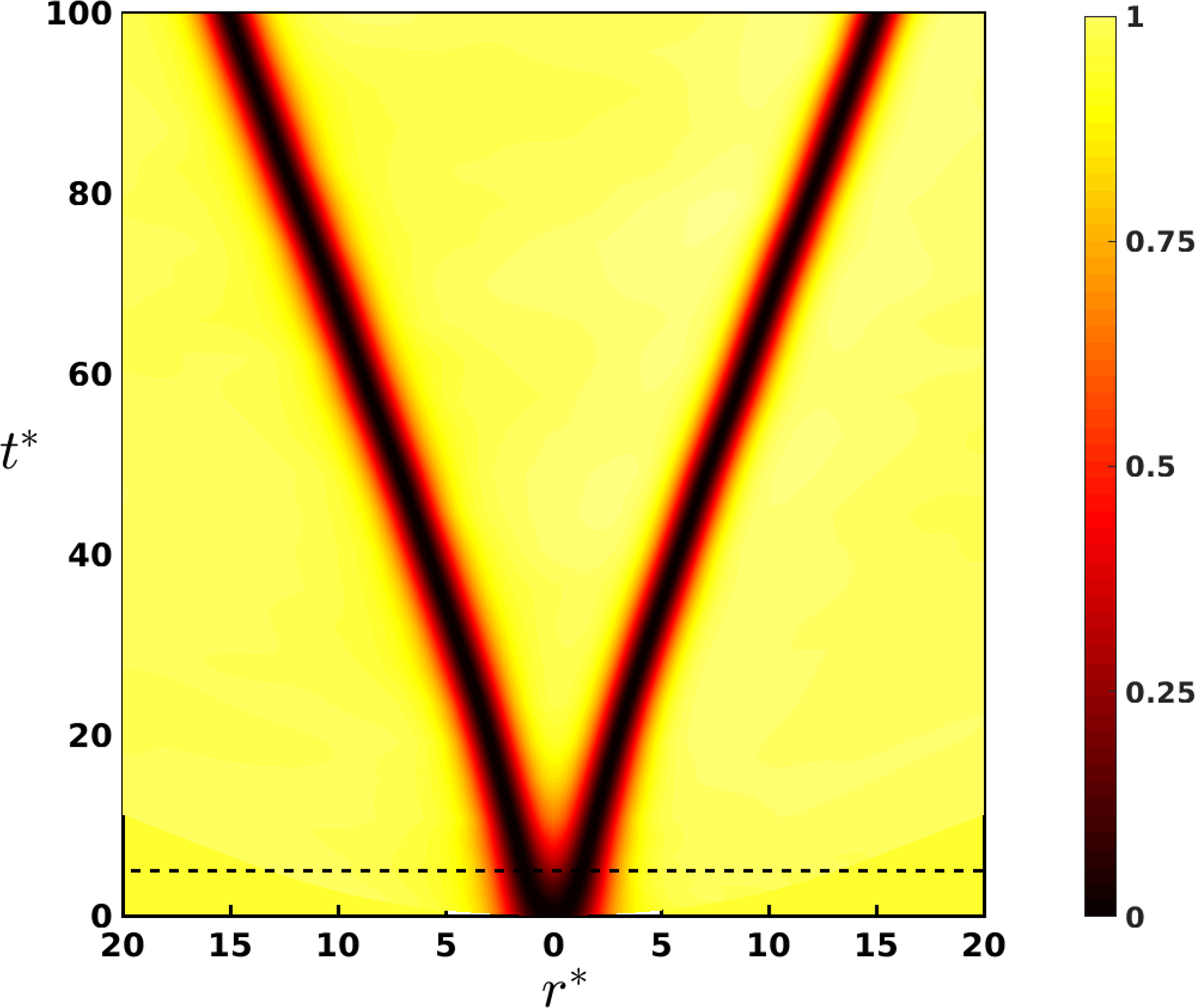}
\caption{{\bf The role of density depletions.} 
Plot of the condensate density along the line containing the 
separation vector $\bm{\delta}$ between the colliding vortices, 
as a function of the distance $r^*$ to the mid-point
of the separation segment and the rescaled temporal 
distance to reconnection $t^*$, for the vortex ring - vortex line 
pre-reconnection dynamics with $R^*_0=5$. In the initial phase of
the approach (top part of the figure), $\delta \sim t^{\ast}$; the
cross-over to the $\delta \sim  t^{\ast^{1/2}}$ scaling occurs when
the vortex cores start to merge (bottom part) for
$t^*\lesssim 5$ (dashed line).}
\label{fig:GP_dens}
\end{figure}

\subsection{The role of density depletions}
Current and previous GP simulations of reconnections in homogeneous and trapped BECs, see a clear
symmetric pre/post-reconnection $t^{\ast^{1/2}}$ scaling in the 
region $\delta^*\lesssim 5$, irrespective of the initial condition.  The effect is robust and mostly went unnoticed, as
the prefactors $a_{_{1/2}}$ in Eq.~(\ref{eq:scal.0.5}) may vary
depending on the conditions and between the approach/separation.

Figure~\ref{fig:GP_dens} shows the condensate density along the line 
containing the separation vector between two reconnecting vortices (taken to be the ring-line scenario in a homogeneous system), 
as a function of $t^*$
and the distance $r^*=r/\xi$ to the mid-point of the separation segment. 
It is clear that 
for $t^*\lesssim 5$, which is when the $t^{*^{1/2}}$ scaling appears,
the density between the two vortices drops dramatically. 
This behaviour is generic - we obtain it also for any vortex reconnection set-up and across homogeneous and trapped BECs.
This result confirms
the analytical work of Nazarenko and West \cite{nazarenko-west-2003}, who
Taylor-expanded the solution of the GP for reconnecting vortex lines and
predicted the observed $t^{1/2}$ scaling in this limit of
vanishing density (in this limit the cubic nonlinear term 
vanishes, reducing the GP equation to the linear Schr\"odinger equation).
There are hence two different arguments for the observation of the 
$t^{1/2}$ scaling:
the interaction-driven dynamics argument, 
underlying the dimensional scaling of Eq. (\ref{eq:dim.1/2}),
and the vanishing density argument from
the GP equation. The arguments
are both valid
at small lengthscales, consistently with the $t^{1/2}$ scaling observed close to reconnection in all GP simulations. 



\section*{Conclusions}

We have addressed the question of whether there is
a universal route to quantum vortex reconnections by 
performing an extensive campaign of numerical simulations
using the two main mathematical models available
(the Gross-Pitaevskii equation and Vortex Filament method).
What distinguishes our work from previous studies is that, firstly, we have studied
the {\it two} main physical systems which display
quantised vorticity (trapped atomic condensates and superfluid helium),
and, secondly, that we have considered the behaviour over distances one order of magnitude
larger.
By applying rigorous dimensional arguments, we have found that
the minimum distance between reconnecting vortex lines may obey
two fundamental scaling law regimes: the already observed
$\delta^{\ast} \sim t^{\ast^{1/2}}$ scaling 
and a new $\delta^{\ast} \sim t^{\ast}$ scaling.

\begin{figure}[htbp]
\centering
\includegraphics[width=0.98\columnwidth]{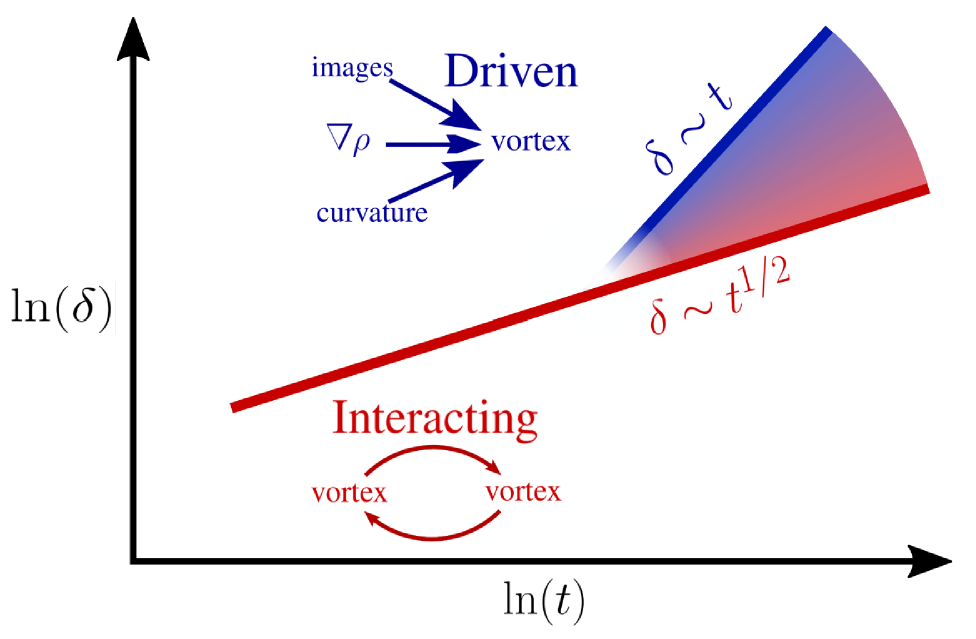}
\caption{
{\bf Fundamental scalings for the reconnection of two vortex lines.}
At small lengthscales the $\delta^{\ast} \sim t^{\ast^{1/2}}$ scaling (in red) is observed as 
the dynamics is determined by the mutual interaction of the two reconnecting vortex strands.
This scaling appears to be universal. At larger distances, we observe two fundamental limiting scenarios:
if the motion is still predominantly driven by the interaction, the $\delta^{\ast} \sim t^{\ast^{1/2}}$ scaling (in red)
still holds; if the dynamics is governed by extrinsic factors driving the individual vortices
a linear $\delta^{\ast} \sim t^{\ast}$ behaviour is established (in blue). In this last case, a scaling crossover occurs.
At large distances, intermediate scalings can arise due to additional physics, {\it e.g.} Kelvin waves (in red-blue color gradient).}
\label{fig:universal}
\end{figure}

At small lengthscales, we always observe the 
$\delta^{\ast} \sim t^{\ast^{1/2}}$ scaling; this arises from either the mutual interaction between reconnecting strands
or from
the depleted density/nonlinearity in the reconnection region. The observation of this scaling 
in all GP simulations, independently
of the precise nature of the system (homogeneous or trapped)
and initial vortex configuration,
adds 
futher evidence for the existence of this universal $\delta^{\ast} \sim t^{\ast^{1/2}}$ scaling law  
close to reconnection.
At larger lengthscales, two fundamental limiting cases 
appear: the continuation of the $\delta^{\ast} \sim t^{\ast^{1/2}}$ scaling if 
the dynamics is still governed by the vortex mutual interaction,
or a linear $\delta^{\ast} \sim t^{\ast}$ behaviour if vortices 
are individually driven 
by extrinsic factors, such as curvature, density gradients and boundaries/images. In the
latter case, the crossover between the two scaling regimes is determined by the balance
between interaction-dominated motion and individually driven dynamics.  
This scaling behaviour is summarised schematically in Fig.~\ref{fig:universal}. 
We stress that these two fundamental scaling laws represent limiting behaviours:
intermediate scalings can arise due to additional physics, {\it e.g.} Kelvin waves. 
We also stress that the $\delta^{\ast} \sim t^{\ast}$ cannot arise from a uniform flow, 
which would simply advect both vortices in the same direction. 
Instead, it arises in distinct systems, both homogeneous and inhomogeneous,
from the different illustrated physical mechanisms, and has not yet been reported in the literature.

While in homogeneous systems the $t^{\ast^{1/2}}$  behaviour can persist to arbitrary separations 
(e.g. for initially orthogonal and weakly-curved vortices), we find that in trapped condensates the scaling crossover always arises.  
Indeed, the current technological ability to directly image vortex lines
in trapped condensates suggests that full 3D reconstructions will soon be
available putting the detection of this crossover within experimental reach.


%

\begin{acknowledgments}
L.G., C.F.B. and N.G.P. acknowledge the support of  
the Engineering and Physical Sciences Research Council 
(Grant No. EP/R005192/1).
\end{acknowledgments}



%






\end{document}


\title{Crossover from interaction to driven regimes\\
in quantum vortex reconnections}

\author{Luca Galantucci}
\email[]{luca.galantucci@newcastle.ac.uk}
\affiliation{Joint Quantum Centre (JQC) Durham--Newcastle, and
School of Mathematics and Statistics, Newcastle University,
Newcastle upon Tyne, NE1 7RU, United Kingdom}

\author{A.~W.~Baggaley}
\affiliation{Joint Quantum Centre (JQC) Durham--Newcastle, and
School of Mathematics and Statistics, Newcastle University,
Newcastle upon Tyne, NE1 7RU, United Kingdom}

\author{N.~G.~Parker}
\affiliation{Joint Quantum Centre (JQC) Durham--Newcastle, and
School of Mathematics and Statistics, Newcastle University,
Newcastle upon Tyne, NE1 7RU, United Kingdom}

\author{C.~F. Barenghi}
\affiliation{Joint Quantum Centre (JQC) Durham--Newcastle, and
School of Mathematics and Statistics, Newcastle University,
Newcastle upon Tyne, NE1 7RU, United Kingdom}

\date{\today}

\maketitle


\centerline{\LARGE {\bf Supporting Information}}
\vskip 1cm

\section{SI.1: Dimensional analysis}

In the main text, we conjectured
that the minimum distance $\delta$ between reconnecting vortex
lines depends upon 
the temporal distance $t$ to reconnection, 
the quantum of circulation $\kappa$, a characteristic 
lengthscale $\ell$ associated to the geometry of the vortex configuration,
the density $\rho$ and 
the density gradient $\nabla\rho$, postulating the existence
of a functional relation $f$ such that
\begin{equation}
\displaystyle 
f (\delta, t, \kappa, \ell, \rho, \nabla\rho) = 0   \; .\label{eq:dim} 
\end{equation}
Following the Buckingham $\pi$-Theorem \cite{buckingham-1914}, 
there are only three independent dimensionless quantities, 
$\pi_1$, $\pi_2$ and $\pi_3$,
which we choose as: $\pi_1=\delta^2/(\kappa t)$, 
$\pi_2=\ell \delta /(\kappa t)$ 
and $\pi_3=\rho \delta/(\kappa t \nabla\rho)$. 
Other dimensionless quantities can be assembled from the same physical 
variables, but they would 
depend on $\pi_1$, $\pi_2$ and $\pi_3$. 

Equation~(\ref{eq:dim}) can be rewritten as $\pi_1=F_1(\pi_2,\pi_3)$, 
or, equivalently, as $\pi_2=F_2(\pi_1,\pi_3)$, or as
$\pi_3=F_3(\pi_1,\pi_2)$. 
Choosing $F_1(\pi_2,\pi_3)=C_1$, $F_2(\pi_1,\pi_3)=C_2$ and 
$F_3(\pi_1,\pi_2)=C_3$ where $C_1$, $C_2$ and $C_3$ are constant,
we find the scalings reported in the main text, namely

\begin{eqnarray}
\displaystyle
\delta(t) &=& (C_1\kappa)^{1/2}t^{1/2} \label{eq:dim.1/2} \;\; , \\[2mm]
\delta(t) &=& C_2 \left ( \frac{\kappa}{\ell} \right ) t \label{eq:dim.ell} \;\;\;\; \text{and}\\[2mm]
\delta(t) &=& C_3 \left (\kappa\frac{\nabla\rho}{\rho} \right ) t \label{eq:dim.rho} \;\; , 
\end{eqnarray}
where $C_1$, $C_2$ and $C_3$ are non-dimensional constants. 
We conclude that there are two distinct scalings:
$\delta \sim t^{1/2}$ and $\delta \sim t$.

\section*{SI.2: Curvature contribution}

To find the velocity contribution
$v_{\gamma}$ arising from the local vortex curvature
to the relative approach velocity of the reconnecting vortex lines,
$d \delta^*/d t^*$, we proceed as follows. 
First we compute the velocities $\mathbf{v}_{\gamma,1}$ 
and $\mathbf{v}_{\gamma,2}$ 
arising from local vortex curvature
of the two points 
$\mathbf{x}_{1}$ and $\mathbf{x}_{2}$ corresponding to the minimum distance. 
We assume that each velocity
can be approximated by the velocity of a circular 
vortex ring \cite{roberts-grant-1971} of radius $R_c$
corresponding to the local curvature $K=1/R_c$ of that vortex strand,
which is obtained by taking suitable derivatives with respect to
arc length:

\begin{equation}
\displaystyle
\label{eq:vortex.curv.formula}
\mathbf{v}_{\gamma,i}=\frac{\kappa}{4\pi R_{c,i}}\left [ \ln \left ( \frac{8 R_{c,i}}{\xi}\right ) - 0.615 \right ]\!\!\hat{\mathbf{b}} \;\; ,
\end{equation}
where $\hat{\mathbf{b}}$ is the unit binormal vector. 
Secondly, we project 
$(\mathbf{v}_{\gamma,2} - \mathbf{v}_{\gamma,1})$ 
on the separation vector 
$\bm{\delta}=\mathbf{x}_{2}-\mathbf{x}_{1}$,
obtaining 
$v_{\gamma}~=~(\mathbf{v}_{\gamma,2}~-~\mathbf{v}_{\gamma,1})\cdot\hat{\bm{\delta}}$. In order not to overestimate the curvature-driven velocity
in the orthogonal reconnection where for most of the simulated dynamics the curvature $K$ is very small ($K=0$ at the start of simulation), in the calculation of  
$\mathbf{v}_{\gamma}$ for the orthogonal configuration (Fig. 4 (top) main manuscript) we omit the logarithmic correction in Eq. (\ref{eq:vortex.curv.formula}).\\[1mm] 

\section*{SI.3: Homogeneous unbounded systems. GP simulations. Ring - line reconnection.}

In this paragraph, we illustrate in more detail the separation (post-reconnection) dynamics of a vortex-ring - vortex-line reconnection occurring in a homogeneous BEC, investigated via GP
simulations in the main manuscript (Fig.~3 (right)). The aim is to account for the late recovery of the linear $t^{\ast^1}$ scaling of the minimum distance $\delta^*$. As described in the main text,
the linear scaling is observed when (a) the dynamics is governed by the individual motion of the vortices and (b) the projection of the relative velocity 
on the separation vector $\bm{\delta}$ is constant with respect to time. In the ring - line reconnections investigated in the present work, the reconnection generates a Kelvin-wave 
perturbed vortex-line and a non-planar quasi-elliptical vortex-ring (hereafter with 'ring' we intend more generically a closed vortex curve). 
Due to the presence of these Kelvin waves on the ring, the
ring 
wobbles, and consequently its separation velocity is {\it not} constant with respect to time\cite{arms-hama-1965,dhanak-debernardinis-1981},
explaining the non observation of the scaling crossover $t^{*^{1/2}}$ to $t^{\ast^1}$
as soon as the vortex cores separate.
The wobbling vortex-ring generates
sound waves transforming 
incompressible kinetic energy of the vortex ring in acoustic (compressible) 
kinetic energy,
decreasing the length of vortex lines 
(which, in the first approximation, is proportional to the
incompressible kinetic energy).

\begin{figure*}[ht]
\centering
\includegraphics[width=0.8\textwidth]{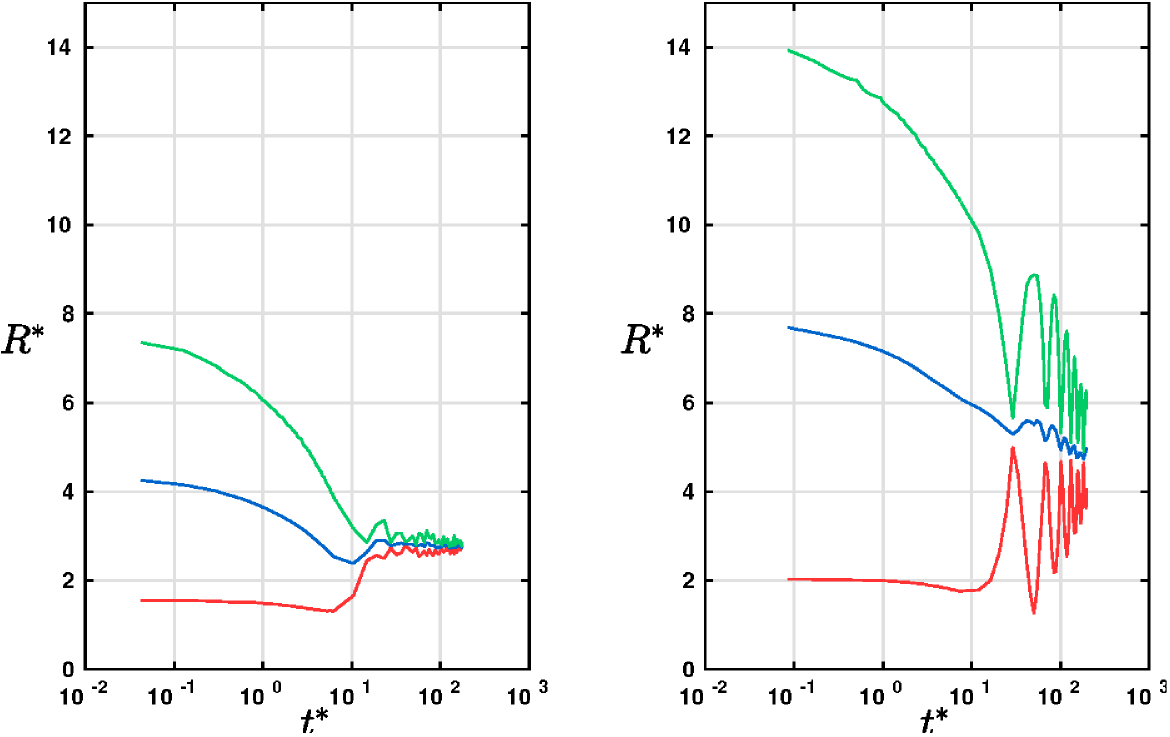}
\caption{{\bf GP simulations: post reconnection vortex-ring wobbling}. 
Post-reconnection temporal evolution of maximum (solid green line), minimum (solid red line) and average 
(solid blue line) semiaxes $R$ of the vortex-ring produced by the ring-line reconnection described in Fig.~3 (right) of main manuscript. 
{\bf Left}: initial vortex-ring radius $R_0^*=5$. {\bf Right}: initial vortex-ring radius $R_0^*=10$. 
} 
\label{fig:Rings_oscill}
\end{figure*}

The oscillation dynamics of the vortex-ring is illustrated 
in Fig.~\ref{fig:Rings_oscill}, where we have reported the temporal evolution of 
the maximum, minimum and average semiaxes of the vortex-ring, after the reconnection with the vortex-line. 
As the closed vortex curve is neither a planar curve or perfectly symmetrical, with `semiaxes' we intend the distance of the points $\mathbf{x}(\zeta)$ lying on the 
vortex curve (where $\zeta$ is arclength) from the vortex geometric centre 
$\displaystyle\mathbf{x}_G=\oint_{_{\rm ring}}\!\!\!\!\!\! \mathbf{x}(\zeta)d\zeta{\bigg /}\!\!\oint_{_{\rm ring}}\!\!\!\!\!\!\!\!d\zeta$.

\noindent
Figure~\ref{fig:Rings_oscill} (left) clearly illustrates that in the simulation with initial vortex-ring radius $R_0^*=5$, the oscillation becomes negligible for $t^*\gtrsim 50$ where the vortex
recovers a circular shape hence travelling at a constant velocity. For $t^*\gtrsim 50$ we indeed recover the linear scaling $\delta^*(t^*)\sim t^*$ (Fig.~3 (right) 
in main manuscript). In contrast, in the simulation with initial radius $R_0^*=10$, at $t^*\approx 200$ the vortex-ring is still wobbling 
(Fig.~\ref{fig:Rings_oscill} (right))
explaining why for these initial conditions the linear scaling is not (so far) recovered.

The recovery of the $t^{*^{1}}$ scaling at large distances in the post-reconnection dynamics confers to the ring - line configuration 
a broad relevance in the context of quantum vortex reconnections, as it highlights its limiting character.

\newpage

\newpage

\section*{SI.4: VFM simulations}
Numerical simulations performed using the VFM (Vortex Filament Method)
provide further evidence for the scaling laws presented in our article.
The VFM \cite{hanninen-baggaley-2014} applies very well to vortex
dynamics in superfluid helium turbulence because of the large
separation of scales typical of such problem: 
$D \gg \ell_v \gg a_0$, where $D$ is the size of
the system, $\ell_v$ the average intervortex distance, and $a_0$ the vortex
core size.
When using the VFM to study reconnections of quantum vortices 
\cite{dewaele-aarts-1994}), we must make the following important caveats. 
Firstly, the VFM does not give information about lengthscales of the order
of or smaller than  $\Delta \zeta$, defined as
the spatial discretisation along the vortex lines; typically,
$\Delta \zeta / a_0 \gtrsim 10^5$. 
Secondly, in the VFM, the motion of 
vortex line elements is governed by the Biot-Savart law, which 
formulates the classical Euler equation in integral form. Since vortex
reconnections are outside the realm of Euler dynamics,
an ad-hoc artificial {\it cut and paste} algorithm must be implemented
\cite{baggaley-2012b} (see Methods for further details). 
Because of the presence of the reconnections algorithm, the VFM cannot
provide physical information at lengthscales smaller than $2 \Delta \zeta$
or $3 \Delta \zeta$.

\begin{figure}[ht]
\centering
\begin{minipage}{0.47\textwidth}
      \centering
       \includegraphics[width=0.95\textwidth]{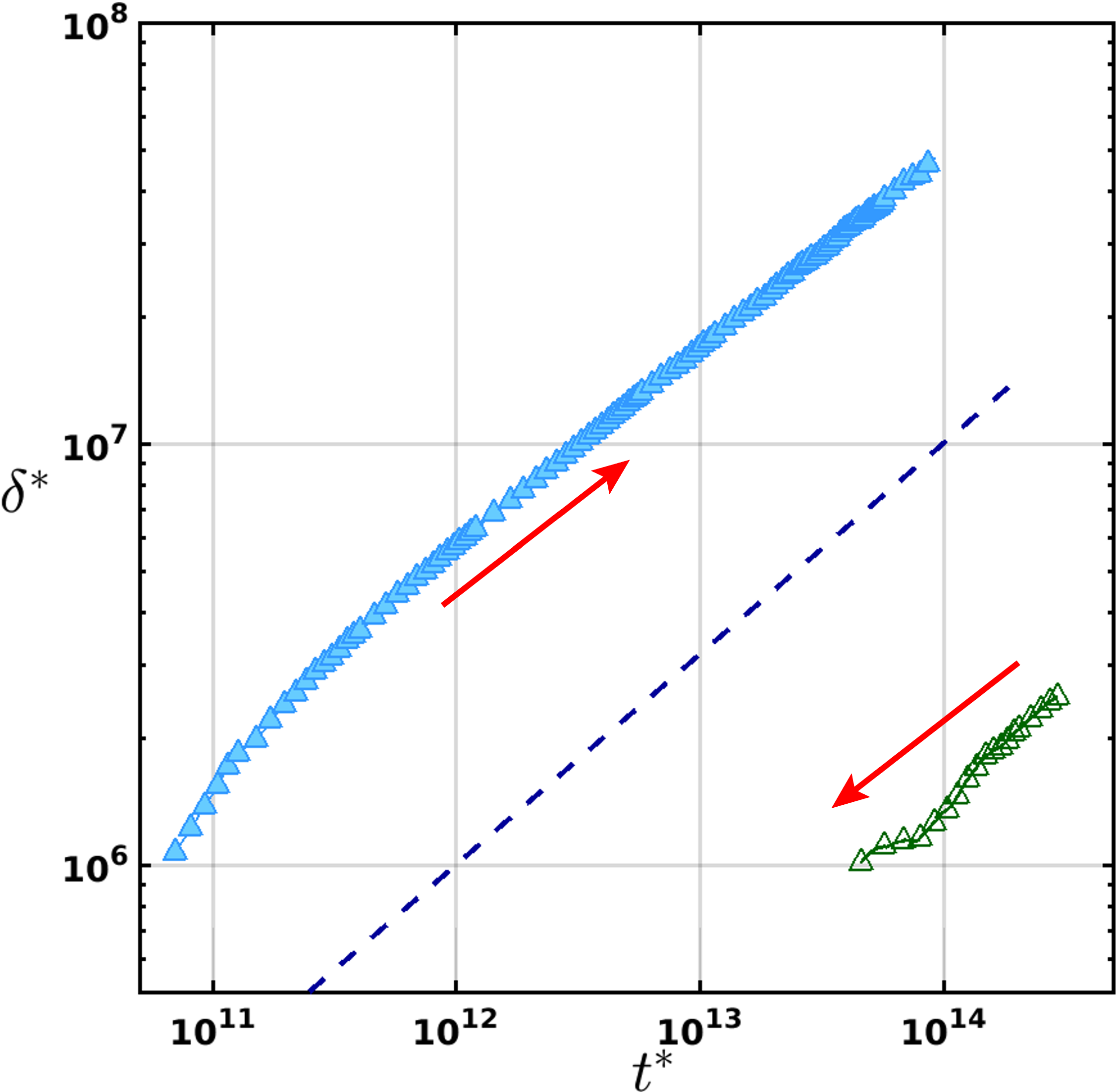}
     \end{minipage}
    \hspace{0.004\textwidth}
     \begin{minipage}{0.47\textwidth}
      \centering
       \includegraphics[width=0.95\textwidth]{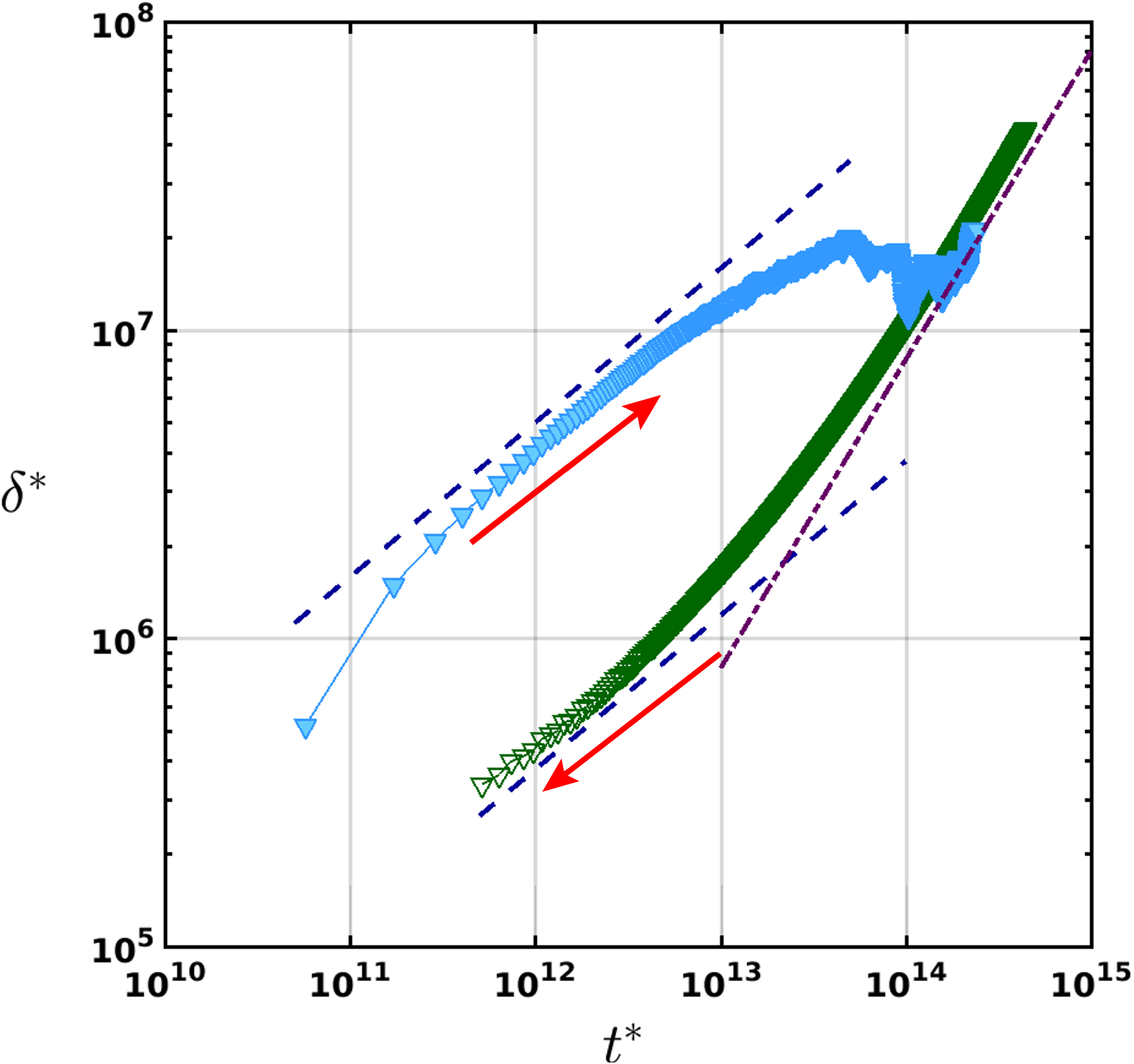}
     \end{minipage}
\caption{{\bf VFM simulations: homogeneous, unbounded and incompressible superfluid $^4$He systems}.
Evolution of the  minimum distance $\delta^*$ between reconnecting vortices as a function of the  temporal distance to reconnection $t^*$. 
Open green (filled cyan) symbols correspond to pre (post) reconnection dynamics. {\bf Left}: reconnection between initially orthogonal vortices with initial distance $\delta^*_0=2.5\times 10^6$.
{\bf Right}: vortex-ring - vortex-line reconnection for initial distance $\delta^*_0=4.5\times 10^7$ and vortex-ring initial radius $R_0^*=4.2\times 10^7$. 
In both subfigures, the blue-dashed line shows the $t^{*^{1/2}}$ scaling. The dot-dashed violet line in the right panel indicates the $t^{\ast}$ scaling. Red arrows indicate the direction of time.} 
\label{fig:VFM}
\end{figure}

Still, information about the minimum vortex distance $\delta(t)$
obtained by the VFM at lengthscales larger than $\Delta \zeta$
is important and supports our arguments.
In Fig.~\ref{fig:VFM} (left), we report the scaling of $\delta$ for 
initially orthogonal vortices, separated by an initial distance
$\delta_0^*=\delta_0/a_0=2.5\times 10^6$. For both pre/post-reconnection 
dynamics, we only observe the $t^{\ast^{1/2}}$ scaling, as expected, as
the motion of the vortex lines is governed by their mutual interaction.


Fig.~\ref{fig:VFM} (right) illustrates the temporal evolution of $\delta^*$
for a ring - line reconnection, and
clearly shows the scaling cross-over 
(arising from the balance between interaction and self-driven motions)
which we observe in the corresponding GP simulations.
The cross-over takes place when $\delta(t) \lesssim R_c(t)$, $R_c(t)$ being the 
radius of curvature of the ring in the point $\mathbf{x}_{\rm ring}$ which at time $t$ is at minimum distance
from the vortex - line. When $\delta(t)\lesssim R_c(t)$, in fact, the mutual interaction is
predominant tending hence towards the $t^{\ast^{1/2}}$ scaling. The smaller cross-over scale 
with respect to the initial vortex ring radius $R_0$ arises from the increasing curvature of the ring while travelling towards the line
as an effect of their interaction, {\it e.g.} at $t_{cr}^*\approx 10^{13}$, in the cross-over region, $R^*_c(t_{cr}^*)\approx 4\times 10^6$
(while $\delta^*(t_{cr}^*)\approx 2\times 10^6$).

It is also important to note that in the VF method, the spatial discretisation
$\Delta \zeta$ on the vortices constrains the smallest radius of 
curvature $R_{c,min}\sim 5\,\Delta \zeta$ which can numerically be resolved.
As a consequence, for $\delta \lesssim R_{c,min}$ the scaling will always tend to 
$t^{\ast^{1/2}}$.

This $t^{\ast^{1/2}}$ to $t^{\ast^{1}}$ crossover in the ring - line reconnection
is best recognized in the pre-reconnection stage
(green open symbols) in Fig.~\ref{fig:VFM} (right), where the curve
smoothly changes slope. The interpretation of the post-reconnection 
behaviour (cyan filled symbols) is complicated by the
role played by Kelvin waves which are excited by the reconnection event and modify the motions of the line and ring.  
The scaling exponent is exactly $1/2$ for $ 2.5\times 10^{11}\lesssim t^* \lesssim 5\times 10^{12}$), 
but we cannot directly ascribe this as being to the mutual interaction of vortices due to the unclear impact of the Kelvin waves.

Movies S13 - S16 show the orthogonal and ring - line reconnections performed with the VFM.

\newpage

\newpage

\section*{SI.5: Box-trapped BECs. GP simulations}

In order to study the scaling of the minimum distance $\delta$ with respect to time in further experimentally accessible geometries and with the objective of backing our
interpretation of the scaling crossover outlined in the main text, we perform numerical simulations of vortex dynamics and reconnections in recently designed {\it box-traps} 
\cite{gaunt-etal-2013,navon-etal-2016}. The fundamental differences between box-trapped BECs and harmonically trapped BECs (investigated in the main text, see Figs.~5 and 6) 
are mainly two.
First, in box-traps the 
density of the confined BEC is constant throughout the sample (exception made for a thin layer near the trap edges whose width is of the order of the healing length), while in harmonically trapped BECs
the condensate density is maximum in the center of the trap and decreases parabolically when moving towards the boundaries of the trap. 
Second, while the role of images in both box- and harmonically trapped BECs 
still remains an open question \cite{fetter-svidzinsky-2001,anglin-2002,khawaja-2005,mason-berloff-fetter-2006,fetter-2009},
in box-traps vortex images with respect to boundaries are believed to be the only source of vortex motion\cite{mason-berloff-fetter-2006}; on the contrary, in harmonic traps 
vortex dynamics is also determined by density gradients (arising from the inhomogeneous trapping potential) and vortex curvature 
\cite{svidzinsky-fetter-2000,fetter-svidzinsky-2001,jackson-etal-1999}.

\begin{figure}[ht]
\centering
\includegraphics[width=0.9\columnwidth]{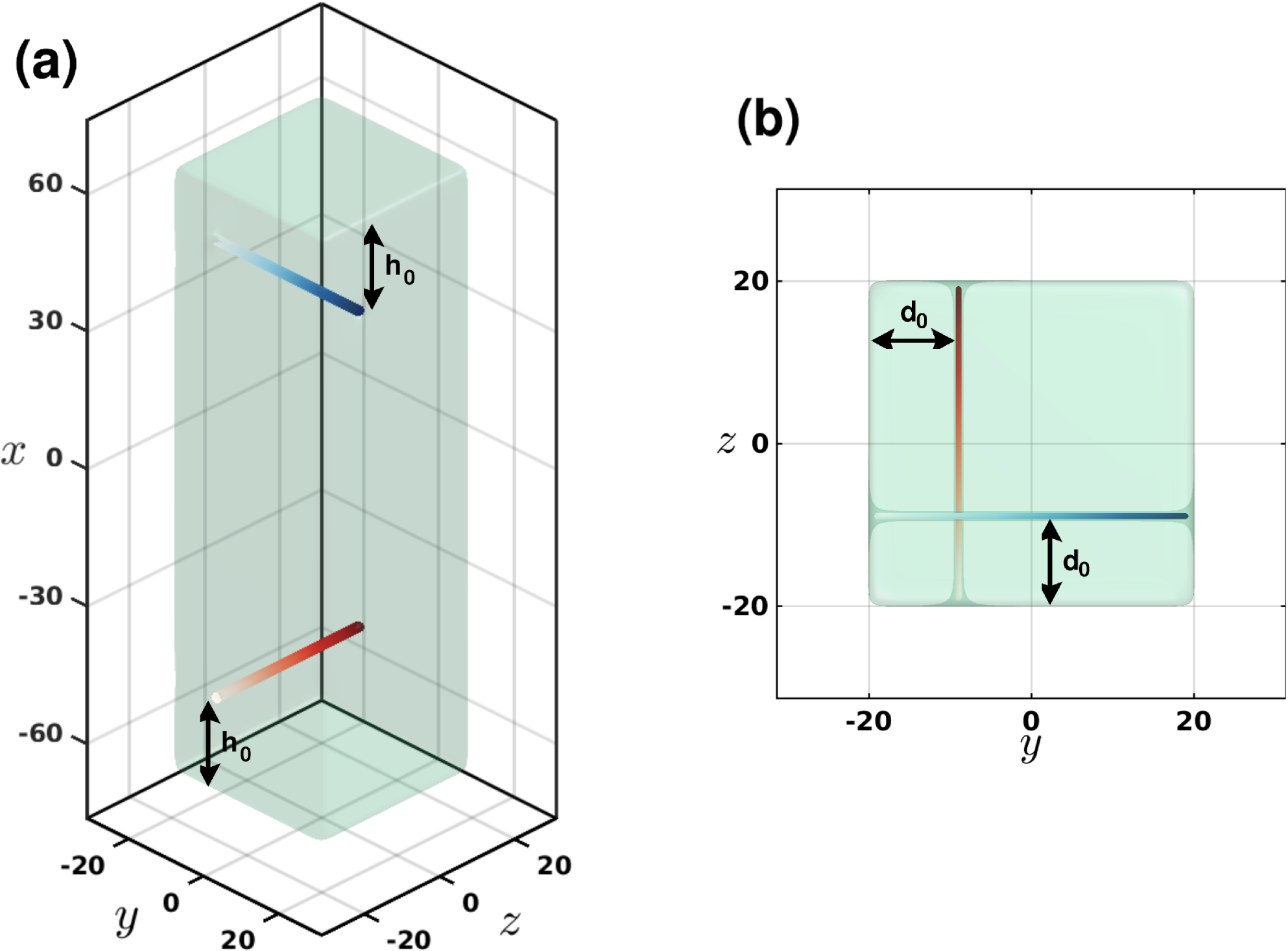}
\caption{{\bf GP simulations: box-trapped BECs, initial conditions}.
{\bf (a), (b)}: Lateral and top view of initial vortex configuration for box trapped BECs; 
Light green surfaces coincide with isosurfaces of condensate density at $5\%$ of trap-center density. 
Color gradient on vortices indicates the direction of the superfluid vorticity (from light to dark).
Unit of length is the healing length $\xi$ evaluated in the bulk of the condensate.} 
\label{fig:traps.ic}
\end{figure}

\begin{figure}[ht]
\centering
\begin{minipage}{0.47\textwidth}
      \centering
       \includegraphics[width=0.95\textwidth]{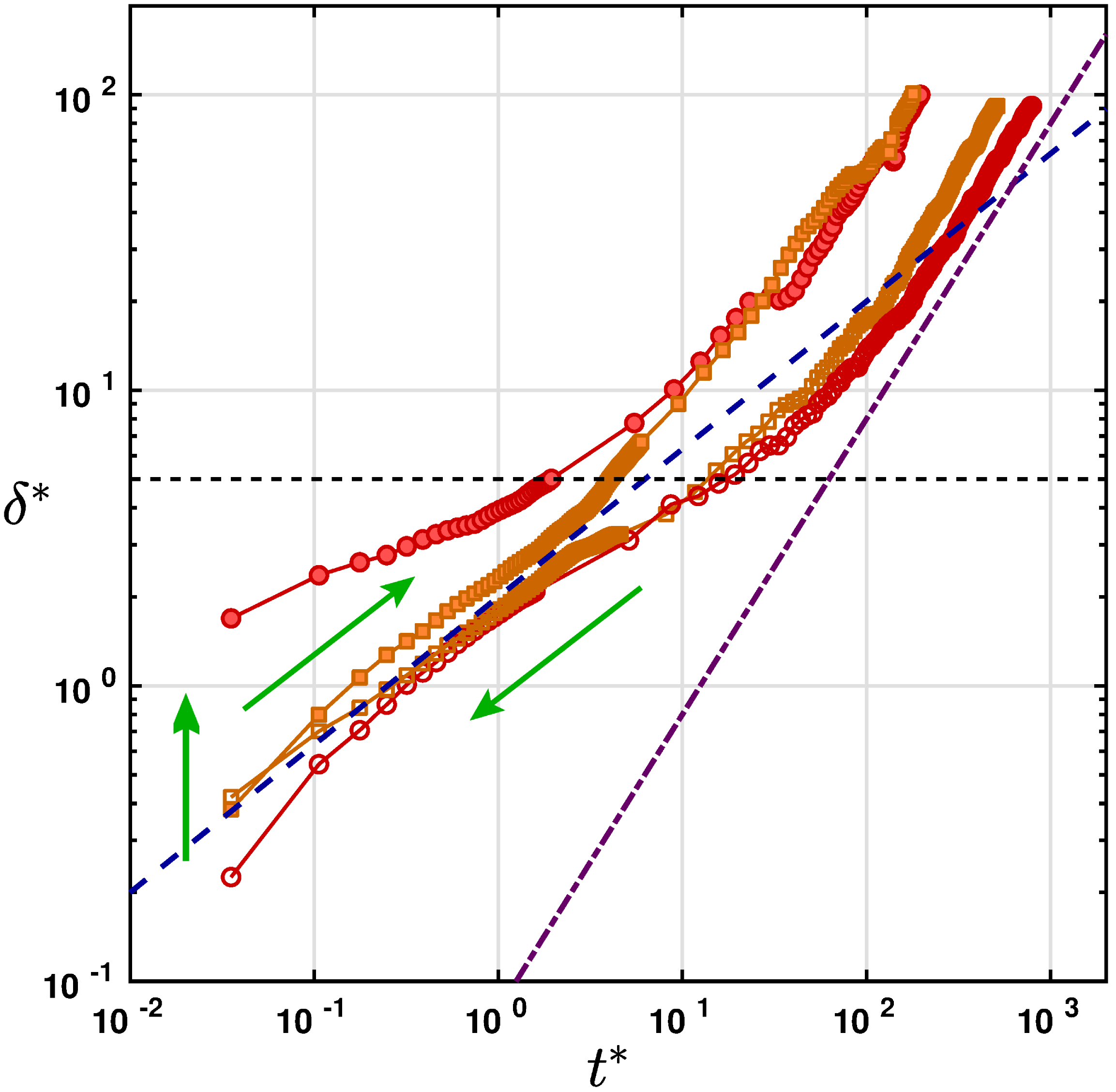}
     \end{minipage}
    \hspace{0.004\textwidth}
     \begin{minipage}{0.47\textwidth}
      \centering
       \includegraphics[width=0.95\textwidth]{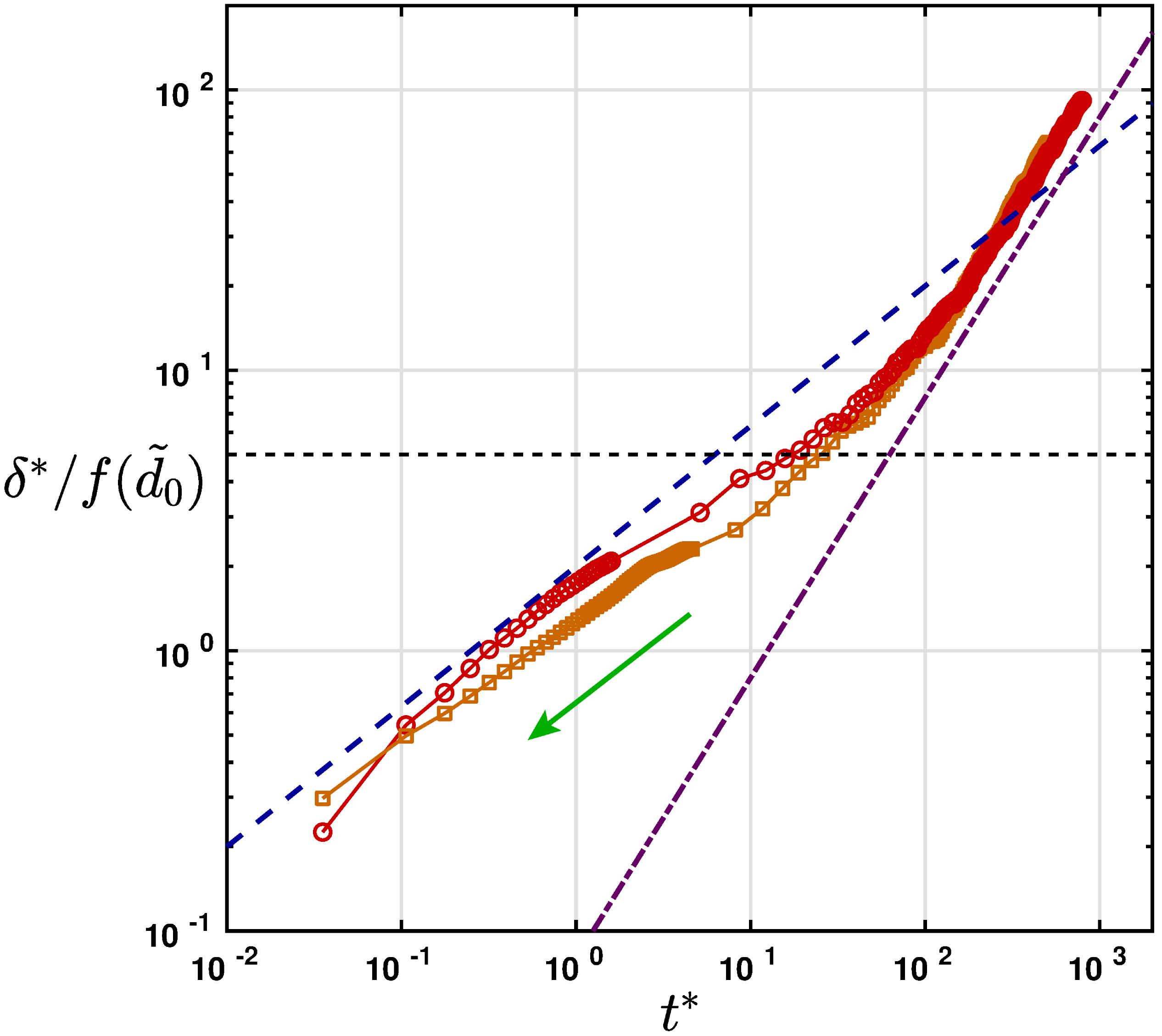}
     \end{minipage}
\caption{{\bf GP simulations: box trapped BECs}.
Evolution of the  minimum distance $\delta^*$ between reconnecting vortices as a function of the temporal distance to reconnection $t^*$. 
Open (filled) symbols correspond to pre (post) reconnection dynamics. {\bf Left}: pre- and post-reconnection scaling of $\delta^*$ for initially imprinted
orthogonal vortices at distance $d_0^*$ from trap lateral boundaries and $h_0^*$ from top/bottom boundaries (see Fig.\ref{fig:traps.ic}):  
$d_0^*=8.25$ (\dyellow{$\bm{\square}$}), $d^*=11.05$ (\dred{$\bm{\circ}$}); $h_0^*=20$ for both simulations. 
{\bf Right}:  pre-reconnection scaling of rescaled minimum distance $\delta^*/f(\tilde{d}_0)$. 
Symbols as in left panel. In both subfigures, the dashed blue and dot-dashed violet lines show the $t^{*^{1/2}}$ and $t^{\ast}$ 
scalings, respectively. The horizontal black dashed line indicates the width of the vortex core ($\approx 5\,\xi$). Green arrows indicate the direction of time.} 
\label{fig:box.traps}
\end{figure}

The initial vortex configuration for the two simulations performed in box-traps is illustrated in Fig.~\ref{fig:traps.ic} (a), (b). The initial distance from the top/bottom boundary $h_0^*=h_0/\xi=20$ is
constant for both simulations (where $\xi$ is the healing length evaluated in the bulk of the condensate, {\it i.e.} away from the trap boundaries), 
while to investigate the potential role of images we choose two distinct values of $d_0^*$: $d_0^*=8.25$ (dark yellow squares in Fig.~\ref{fig:box.traps}) 
and $d_0^*=11.05$ (red circles in Fig.~\ref{fig:box.traps}). 
The temporal evolution of the minimum distance
$\delta^*=\delta/\xi$ is reported in Fig.~\ref{fig:box.traps} (left). It clearly emerges that for $t^* \lesssim 10$, $\delta^*$ follows a $t^{\ast^{1/2}}$ scaling indicating that when vortex cores start merging, 
the predominant dynamics driving the approach of the vortices is the interaction between the latter, 
consistently with the characteristics of vortex reconnections observed in homogeneous condensates (see Fig.~3 in main manuscript).

In addition, for $t^* \gtrsim 200$ in the approach dynamics, 
we again observe a linear scaling, {\it i.e.} $\delta^*\!(t^*) \sim a_{_1} t^{\ast}$, as in homogeneous BECs (see Fig.~3 in main manuscript), implying a constant
value of $d \delta^*/d t^*$.
In the present geometry, we conjecture that this constant projection of the relative velocity between vortices on the separation vector ({\it i.e.} $d \delta^*/d t^*$) coincides with the sum of the velocity magnitudes
induced by the vortex images \cite{mason-berloff-fetter-2006},

\begin{eqnarray}
\displaystyle \frac{d \delta^*}{d t^*} & = & \big | \mathbf{v}_2 - \mathbf{v}_1 \big | = v_{2,img}^* + v_{1,img}^* \nonumber \\[1mm]
& = &  \frac{\kappa}{2 \pi c (d_0 - \sqrt{2}\xi)} = \frac{\kappa}{2 \pi c \overline{d_0}} \, \frac{1}{\left [ \widetilde{d}_0 - \displaystyle\frac{\sqrt{2}}{\overline{d_0}^*} \right ]}\nonumber \\[1mm]
& = & C f(\widetilde{d}_0) \,\,\, \label{eq:delta.der}, 
\end{eqnarray} 

\noindent
where: $\mathbf{v}_1$ ($\mathbf{v}_2$) is the velocity of the top (bottom) vortex; $v_{1,img}$ ($v_{2,img}$) is the velocity of the top (bottom) vortex induced by the corresponding image
with respect to the lateral boundary of the trap at a distance $d_0$ (cfr. Fig.~\ref{fig:traps.ic} (a), (b)); 
$\widetilde{d}_0=d_0/\overline{d_0}$,
$\;\overline{d_0}=9.65\xi$ being the average distance to the lateral boundary of the trap amongst the two
simulations performed in box-traps; $\xi$ is the healing length based on the bulk density $n=\mu/g$, with $\mu$ and $g$ being the chemical potential and the repulsive interaction strength, respectively;
$c=\sqrt{g n/m}$ is the speed of sound;
$\displaystyle C=\frac{\kappa}{2 \pi c \overline{d_0}}$ is a non-dimensional constant independent of $d_0$ and 
$\displaystyle f(\widetilde{d}_0)=\left [\widetilde{d}_0 - \displaystyle\frac{\sqrt{2}}{\overline{d_0}^*} \right ]^{-1}$ is a non-dimensional function depending on the distance $d_0$.

Hence, for $t^* \gtrsim 200$ in pre-reconnection dynamics we argue that $\delta^*\!(t^*) \sim C f(\widetilde{d}_0) t^{\ast}$. This conjecture is confirmed in Fig.~\ref{fig:box.traps} (right)
where we report the temporal evolution of the rescaled minimum distance $\delta^*/f(\widetilde{d}_0)$ for $d_0^*=8.25$ and $d_0^*=11.05$ : the curves
do indeed overlap at large distances. 
This results confirms, for the first time in literature, that quantum vortices, when confined in three-dimensional 
box-trapped condensates, are definitely driven by their images with respect to the boundaries of the condensate.
The post-reconnection dynamics is less straightforward to investigate as a result of other factors playing a significant role in determining the vortex dynamics, above all
acoustic emission and vortex pinning on the boundaries of the trap (as it can be observed in movies M4).\\[1mm] 

\newpage

\newpage

\newpage

\newpage

\section*{Methods}

\subsection*{SI.6: Numerical methods for GP simulations}

Real-time dynamical simulations of vortex dynamics in BECs 
at temperature $T=0$ are performed by numerically integrating the
mean-field Gross-Pitaevskii (GP) equation,

\begin{equation}
i \hbar \frac{\partial \Psi}{\partial t} = - \frac{\hbar^2}{2 m} \nabla^2 \Psi +  V \Psi + g | \Psi |^2 \Psi
\end{equation}
for the complex macroscopic wave function $\Psi(x,y,z,t)$,
where $m$ is the boson mass, 
$V$ is the external potential and $g=4 \pi \hbar^2 a_s/m$ is the strength of the repulsive two-body interaction,
$a_s$ being the atomic scattering length. 

Our code uses second-order accurate finite differences in space
and a fourth order Runge--Kutta method in time.

The algorithm for vortex tracking is based on 
the pseudo-vorticity unit vector 
\begin{equation}
\hat{\bm{\omega}}~:=~\frac{\nabla \Psi_\Re \times \nabla \Psi_\Im}{\left | \nabla \Psi_\Re \times \nabla \Psi_\Im \right |}
\nonumber
\end{equation}
which is tangent to the vortex line along its 
length~\cite{rorai-etal-2016,villois-etal-2017}, where
$\Psi=\Psi_\Re + i\Psi_\Im$. 
\\[1mm]

\noindent
{\bf Homogeneous BECs.}
In homogeneous systems, the external potential $V$ is zero and the wave 
function $\Psi$ is normalized by $\Psi_0=\sqrt{n}$, 
where $n=\mu/g$ is the homogeneous density
of an unbounded BEC in its ground state. By writing
$\Psi^*=\Psi/\Psi_0$ and introducing the healing length
$\xi=\hbar/\sqrt{2m g n}$ as the unit of length
and $\tau=\xi/c$ as the unit of time,
where $c=\sqrt{gn/m}$ is the velocity of sound, we obtain the following 
non-dimensional GP equation:
\begin{equation}
i \frac{\partial \Psi^*}{\partial t^*} = - \frac{1}{2} \nabla^{*^2} \Psi^* + \frac{1}{2} | \Psi^* |^2 \Psi^*
\end{equation} 
where the superscript `$^*$' indicates non-dimensional variables. Hereafter all the quantities mentioned are dimensionless, unless otherwise stated,
and the superscript `$^*$' is omitted to ease notation.\\

\noindent
{\it Orthogonal vortices.}
We imprint two orthogonal vortices employing Pad\'{e} approximants \cite{berloff-2004} for the density field and letting the 
system evolve in imaginary time: the first is
oriented in the negative $z$ direction, the second in the negative 
$y$ direction. The vortices
intersect the $x$ axis at $(-x_0,0,0)$ and $(x_0,0,0)$ respectively.
We compare runs with 
$x_0=5,\, 10,\, 15$  leading to initial distances
$\delta_0=10,\, 20,\, 30$ (see the bottom inset in Fig.~3 (left) of
the main manuscript). Spatial and time discretization use steps
$\Delta x=\Delta y=\Delta z=0.35$ and  $\Delta t = 0.02$. 
The number of grid-points
in the $x$, $y$ and $z$ direction are \{$N_x,N_y,N_z$\}=\{400,400,400\} respectively, leading to a computational box 
$\left \{ \left [ x_{\rm min} : x_{\rm max} \right ] \times \left [ y_{\rm min} : y_{\rm max} \right ] \times \left [ z_{\rm min} : z_{\rm max} \right ] \right \} = 
\left \{ \left [ -70 : 70 \right ] \times  \left [ -70 : 70 \right ] \times \left [ -70 : 70 \right ] \right \}$.\\[1mm]

\noindent
{\it Ring - line reconnection.}
We perform simulations for a vortex ring of radius 
$R_0=5, 7.5$ and $10$,
and centre $(x_0,\, y_0,\, z_0)=(-100,\, 21, \, 0)$ 
on a plane perpendicular to the $x$ direction. The line is
oriented in the positive $z$ 
direction and intersects the $y$ axis at
 $(x_1,\, y_1,\, z_1)=(0,\, 21, \, 0)$,  as shown in the
bottom inset of Fig.~3 (right) in the main manuscript.

The spatial and temporal resolution coincides with 
orthogonal vortices simulations. The numbers of grid-points
in the $x$, $y$ and $z$ directions are 
$N_x=800$, $N_y=280$ and $N_z=160$
respectively, leading to a larger computational box:
$-140 \le x \le 140$, $-50 \le y \le 50$ and $-28 \le z \le 28$.
\bigskip


\noindent
{\bf Trapped BECs.}
For harmonically trapped, cigar-shaped BECs
the units of length, time and energy are $\ell_{tr}=\sqrt{\hbar/(m \omega_\perp)}$, $\tau=\omega_\perp^{-1}$ and $\epsilon=\hbar \omega_\perp$, respectively,
where $\omega_\perp$ is the radial trapping frequency of the harmonic 
potential. The resulting non-dimensional GP equations is

\begin{equation}
i \frac{\partial \widetilde{\Psi}}{\partial \tilde{t}} = - \frac{1}{2} \widetilde{\nabla}^{2} \widetilde{\Psi} + \widetilde{V}\widetilde{\Psi} + \tilde{g} | \widetilde{\Psi} |^2 \widetilde{\Psi}
\end{equation}
where with $\widetilde{\cdot}$ we indicate non-dimensional quantities, 
$\widetilde{\Psi}$ is normalized to unity
($\int_V \vert \widetilde{\Psi} \vert^2 dV=1$ where $V$ is the volume), 
and $\tilde{g}=4 \pi N a_s/\ell_{tr}$ and $N$ is the number of atoms in
the condensates.  Hereafter all the quantities which we mention are meant to be dimensionless unless otherwise stated,
and the superscript $\widetilde{\cdot}$ is omitted to ease notation.\\

\noindent
{\it Box Traps.}
We choose $g=2.35\times 10^4$ and set the trapping potential to
\begin{equation}
\displaystyle
V = \Bigg \{ 
\begin{array}{c}
0 \;\; \text{for} \;\; |x| < L_x/2 \;\; \text{and} \;\; |y| < L_y/2 \;\; \text{and} \;\; |z| < L_z/2 \\[4mm]
10 \mu \;\; \text{elsewhere}
\end{array}
\end{equation}

\noindent
where $\mu=10$ is the chemical potential, $L_x=29.46\ell_{tr}=131.75\xi$ and $L_y=L_z=8.944\ell_{tr}=40\xi$, $\xi$ being the healing length in the bulk of the condensate.
\medskip

In Fig.~\ref{fig:traps.ic} (a), (b), the distance of vortices 
from the top/bottom edge of the trap is $h_0=4.48\ell_{tr}=20\xi$, 
leading to the
initial minimum distance between vortices $\delta_0=20.5\ell_{tr}=91.75\xi$. 
The distance of the vortices from the lateral edges is 
$d_0=1.48\ell_{tr}=8.25\xi$ and $d_0=2.47\ell_{tr}=11.05\xi$. 
The top vortex is oriented in positive $y$ direction and the bottom vortex is the positive $z$ direction.

The grid-spacings are homogeneous in the 
three Cartesian directions,
$\Delta x=\Delta y=\Delta z=7.5\times 10^{-2}$, and the time step 
is $\Delta t = 1.25\times 10^{-3}$. The numbers of grid-points
in the $x$, $y$ and $z$ directions are 
$N_x=512$, $N_y=192$ and $N_z=192$, leading to the
computational box 
$-19.2 \le x \le 19.2$, $-7.2 \le y \le 7.2$ and $-7.2 \le z \le 7.2$.

\medskip

\noindent
{\it Harmonic Traps.}
We choose $g=7.4\times 10^3$ and $\mu=10$. The trapping potential is
\begin{equation}
\displaystyle
V = \frac{1}{2} \left [ \left (\frac{\omega_x}{\omega_\perp}\right )^2 x^2 + r_\perp^2  \right ]
\end{equation}
where $\omega_x=2\pi \times 26 \rm~Hz$ is the axial trapping frequency, 
$\omega_\perp=\omega_y=\omega_z=2 \pi \times 131 \rm~Hz$ 
is the radial trapping frequency, and 
$r_\perp= (y^2 + z^2)^{1/2}$ is the distance from the axis of the 
condensate. These parameters correspond to $R_\perp / \xi_c =2\mu/(\hbar \omega_\perp)=20$,
where $R_\perp$ is the radial Thomas-Fermi radius and $\xi_c$ is the healing length evaluated in the center of the trap.

We imprint two orthogonal vortices as shown
in Fig.~5 of main text employing Pad\'{e} approximants \cite{berloff-2004} for the density field and letting the 
system evolve in imaginary time; the top vortex
is oriented in the negative $y$ direction,
the bottom in the negative $z$ direction. The vortices intersect 
the $x$ axis at $(\pm x_0,0,0)$, where 
$x_0=\chi R_x$, $\chi=0.35,\, 0.5,\, 0.6$
is the orbit parameter, and $R_x$ is the axial Thomas-Fermi radius.

Spatial and temporal resolutions coincide with the box-trap simulations. 
The numbers of grid-points
in the $x$, $y$ and $z$ directions are 
$N_x=800$, $N_y=224$ and $N_z=224$, leading to the
computational box $-30 \le x \le 30$, $-8.4 \le y \le 8.4$, and
$-8,4 \le z \le 8.4$.
\bigskip


\subsection*{SI.7: Numerical methods for VF simulations}

The Vortex Filament method \cite{schwarz-1988,hanninen-baggaley-2014} is a 
well-established model widely employed for the numerical simulations 
of superfluid helium quantum turbulence.
A typical superfluid helium 
experiment is characterised by the large separation of scales between the 
large scale $D$ of the flow ($D\approx 10^{-2}\rm m$) and the small
length scale of the average inter-vortex distance
$\ell_v\approx 10^{-5} \rm m$;  the 
vortex core radius, $a_0\approx 10^{-10}\rm m$ in $^4$He and about 100
times larger in $^3$He-B, is even smaller.
It is therefore appropriate to mathematically model quantum vortices 
as closed space curves of infinitesimal thickness $\mathbf{s}(\zeta,t)$, 
where $\zeta$ is
arclength and $t$ time, moving in an inviscid Euler fluid according to the 
Biot–Savart law \cite{schwarz-1985,saffman-1992},
\begin{equation}
\displaystyle
\frac{\partial \mathbf{s}}{\partial t} = 
\frac{\kappa}{4 \pi}\oint_{\mathcal{L}}\frac{\mathbf{s}'(\eta,t)\times \left [ \mathbf{s}(\zeta,t) - \mathbf{s}(\eta,t) \right ]}{\left | \mathbf{s}(\zeta,t) - \mathbf{s}(\eta,t) \right |^3} d\eta
\label{eq:Biot-Savart}
\end{equation} 

\noindent
where the line integration is performed along the entire vortex configuration $\mathcal{L}$ and $\displaystyle \mathbf{s}'(\zeta,t)=\frac{\partial \mathbf{s}}{\partial \zeta}$ is the 
unit tangent vector.

Vortex-lines are discretised in sets of points $\mathbf{s}_i(t)=\mathbf{s}(\zeta_i,t)$, $i=1\dots N$, 
with initial spatial discretisation $\Delta \zeta = 5\times 10^{-5} \rm m$ for the orthogonal
reconnection and and $\Delta \zeta = 2.5\times 10^{-5} \rm m$ for the ring - line reconnection. The Lagrangian dynamics 
of the vortex points is hence obtained evaluating the discretised Biot-Savart integral starting from a given initial vortex configuration. 
The singularity of the Biot-Savart integral for $\eta \to \zeta$
is fixed by taking into account the finite size of the vortex core,
yielding \cite{arms-hama-1965,schwarz-1985} the following decomposition
between local and non-local contributions: 

\begin{eqnarray}
\displaystyle
\frac{\partial \mathbf{s}}{\partial t}
&=& \frac{\kappa}{4 \pi} \mathbf{s}'(\zeta)\times \mathbf{s}''(\zeta) \ln \left ( \frac{2\sqrt{l_+ l_-}}{e^{1/2}a_0}\right ) +\nonumber \\[1mm]
&+& \frac{\kappa}{4 \pi}\oint_{\mathcal{L'}}\frac{\mathbf{s}'(\eta)\times \left [ \mathbf{s}(\zeta) - \mathbf{s}(\eta) \right ]}{\left | \mathbf{s}(\zeta) - \mathbf{s}(\eta) \right |^3} d\eta \, \, ,
\label{eq:Biot-Savart.desing}
\end{eqnarray}

\noindent
where time dependence has been omitted to ease notation, 
$\mathbf{s}''$ is the normal vector at $\mathbf{s}(\zeta)$ ($|\mathbf{s}''|=K$, $K$ being the curavture at $\mathbf{s}(\zeta)$), 
$l_\pm$ are the lengths of the line segments connected to $\mathbf{s}(\zeta)$
after discretisation, and the integral is evaluated on $\mathcal{L'}$, 
the original vortex configuration $\mathcal{L}$ without the line segments 
adjacent to  $\mathbf{s}(\zeta)$. 

Reconnections are not intrinsically predicted by the VF method, 
as Euler inviscid dynamics forbids such changes of topology.
As a consequence, an additional algorithm has to be employed
which changes the topology of the vortex configuration
when two vortex lines become closer than an arbitrary threshold distance
 which in this work we set to $\Delta \zeta/2$. Moreover,
in order to model the dissipative nature (phonon emission) of 
reconnecting events \cite{leadbeater-etal-2001}, 
the numerical {\it cut and paste} reconnection algorithm 
reduces vortex length (which is taken as
a proxy for the kinetic energy of the vortices). 
It is important to stress, in fact, that given the lengthscales of 
the flow investigated 
(much larger than the characteristic lengthscale of density 
variations in superfluid helium), the VF method is an incompressible model. 
Several reconnections algorithms have been introduced 
in VF method literature. Importantly, 
a recent analysis \cite{baggaley-2012b} has showed
that all these algorithms
produce very similar results, at least in the context of superfluid turbulence.

The number of discretization points, $N$, changes with time as the
simulation progresses (discretization points are
introduced or removed to maintain the numerical resolution along the 
vortex filaments.) In the present study, if the separation between two 
discretization points
becomes greater than $\Delta \zeta$, a new intermediate point is inserted 
with the constraint of preserving the vortex curvature. Similarly,
if the separation becomes less than  $\Delta \zeta/2$, points are removed 
in order to ensure that our shortest scale which is numerically
resolved does not change \cite{baggaley-barenghi-2011c}.

In the present work, time integration is performed employing a third order Runge--Kutta method with time step $\Delta t= 5\times 10^{-4} \rm s$, while 
all spatial derivatives are approximated using fourth-order finite difference schemes which account for varying mesh sizes along the vortex filaments \cite{gamet-etal-1999,baggaley-barenghi-2011c}.

\subsection*{Movies}

\subsubsection*{Movies S1 - S4}
Movies S1, S2, S3 and S4 are a rendering of the reconnection of orthogonal vortices with initial minimum distance $\delta_0^*=10$, 
computed employing the Gross-Pitaevskii (GP) model. 

\subsubsection*{Movies S5 - S8}
Movies S5, S6, S7 and S8 are a rendering of the reconnection between a vortex ring and a vortex line with initial minimum distance $\delta_0^*=100$
and initial vortex ring radius $R_0^*=5$, computed employing the GP model. 

\subsubsection*{Movies S9, S10}
Movies S9 and S10 are a rendering of the reconnection between vortex lines in a harmonically trapped Bose-Einstein condensate
corresponding to an orbit parameter $\chi=0.35$, computed employing the GP model.

\subsubsection*{Movies S11, S12}
Movies S11 and S12 are a rendering of the reconnection between vortex lines in a box-trapped Bose-Einstein condensate
corresponding to an initial lateral distance from the trap boundary $d_0^*=11.05$, computed employing the GP model.
 
\subsubsection*{Movies S13, S14}
Movies S13 and S14 are a rendering of the reconnection of orthogonal vortices with initial minimum distance $\delta_0^*=2.5\times 10^6$, 
computed employing the Vortex Filament (VF) method. 

\subsubsection*{Movies S15, S16}
Movies S15 and S16 are a rendering of the reconnection between a vortex ring and a vortex line with initial minimum distance $\delta_0^*=2\times 10^7$
and initial vortex ring radius $R_0^*=1.2\times 10^7$, computed employing the VF method.
